\documentclass[usenatbib, useAMS, fleqn]{mn2e}
\usepackage{mathtools}
\usepackage{amssymb}
\usepackage{gensymb}
\usepackage{upgreek}
\usepackage{epsfig, graphicx}
\usepackage[caption=false]{subfig}
\usepackage{times}

\newcommand{\abs}[1]{\left|#1\right|}
\newcommand{\ud}{\mathrm{d}}
\newcommand{\Msun}{\ensuremath{\mathrm{M}_{\odot}}}
\newcommand{\Mbh}{\ensuremath{M_\bullet}}

\newcommand{\apjl}{ApJL}
\newcommand{\aj}{AJ}
\newcommand{\apj}{ApJ}
\newcommand{\mnras}{MNRAS}

\newcommand{\apjs}{ApJS}
\newcommand{\araa}{ARAA}
\newcommand{\aap}{A\&A}
\newcommand{\nat}{Nature}
\newcommand{\pr}{Phys.~Rep.}
\newcommand{\pasj}{PASJ}

\title[Fate of SNRs near SMBHs]{The fate of supernova remnants near quiescent supermassive\\black holes}

\author[Rimoldi et al.]{A.~Rimoldi,$^{1}$\thanks{E-mail: rimoldi@strw.leidenuniv.nl} E.~M.~Rossi,$^{1}$ T.~Piran$^2$ and S.~Portegies Zwart$^{1}$ \\
$^1$Leiden Observatory, Leiden University, PO Box 9513 Leiden, NL-2300 RA, The Netherlands\\
$^2$Racah Institute of Physics, The Hebrew University, Jerusalem 91904, Israel}
\date{Accepted 2014 December 9. Received 2014 November 27; in original form 2014 October 10}
\pagerange{\pageref{firstpage}--\pageref{lastpage}} \pubyear{2014}

\begin{document} 

\maketitle
\label{firstpage}

\begin{abstract}
There is mounting observational evidence that most galactic nuclei host both supermassive black holes (SMBHs) and young populations of stars.
With an abundance of massive stars, core-collapse supernovae are expected in SMBH spheres of influence.
We develop a novel numerical method, based on the Kompaneets approximation, to trace supernova remnant (SNR) evolution in these hostile environments, where radial gas gradients and SMBH tides are present.
We trace the adiabatic evolution of the SNR shock until $50\%$ of the remnant is either in the radiative phase or is slowed down below the SMBH Keplerian velocity and is sheared apart.
In this way, we obtain shapes and lifetimes of SNRs as a function of the explosion distance from the SMBH, the gas density profile and the SMBH mass. 
As an application, we focus here exclusively on quiescent SMBHs, because their light may not hamper detections of SNRs and because we can take advantage of the unsurpassed detailed observations of our Galactic Centre.
Assuming that properties such as gas and stellar content scale appropriately with the SMBH mass, we study SNR evolution around other quiescent SMBHs.
We find that, for SMBH masses over $\sim 10^7~\Msun$, tidal disruption of SNRs can occur at less than $10^4~\mathrm{yr}$, leading to a shortened X-ray emitting adiabatic phase, and to no radiative phase.
On the other hand, only modest disruption is expected in our Galactic Centre for SNRs in their X-ray stage.
This is in accordance with estimates of the lifetime of the Sgr A East SNR, which leads us to expect one supernova per $10^4~\mathrm{yr}$ in the sphere of influence of Sgr A*.
\end{abstract}


\section{Introduction} \label{sec:introduction}
There is compelling evidence for a supermassive black hole (SMBH) with a mass of $4.3 \times 10^{6}~\Msun$ in the nucleus of the Milky Way, associated with the Sgr A* radio source.
The strongest evidence comes from the analysis of orbits of the so-called `S-stars' very near this compact object, such as that of the star S2 with a period of only $16~\mathrm{yr}$ and pericentre of $\sim 10^2~\mathrm{au}$ \citep{Schodel02, Schodel03, Ghez03, Eisenhauer05, Ghez08, Gillessen09}.

Most other massive galaxies contain SMBHs \citep{Marleau13}, some with masses as high as $10^{10}~\Msun$ \citep{McConnell11}.
The observed fraction of active nuclei is no more than a few per cent at low redshifts \citep{Schawinski10}, and most galactic nuclei house very sub-Eddington SMBHs, like Sgr A* \citep{Melia01, Alexander05, Genzel10}.
These SMBHs are believed to be surrounded by radiatively inefficient accretion flows (RIAFs), where only a small fraction of the accretion energy is carried away by radiation \citep{Ichimaru77, Rees82, Narayan94}.

In addition to the ubiquity of SMBHs, young stellar populations and appreciable star formation rates are common in many quiescent galactic nuclei \citep{Sarzi04, Walcher06, Schruba11, Kennicutt12, Neumayer12}.\footnote{Evidence for recent star formation has also been seen around active galactic nuclei \cite[AGN; for example,][]{Davies07}. However, active nuclei are not the subject of this study.}
This is seen most clearly in the abundance of early-type stars in the central parsec of the Milky Way \citep[see][for some recent reviews]{Do13a, Do13b, Lu13}.
Moreover, it appears that star formation in the Galactic Centre region has been a persistent process that has increased over the past $10^8~\mathrm{yr}$ \citep{Figer04, Figer09, Pfuhl11}.
Over that time, an  estimated $\gtrsim 3 \times 10^5~\Msun$ of stars have formed within $2.5~\mathrm{pc}$ of the SMBH \citep{Blum03, Pfuhl11}.

Continuous star formation in galactic nuclei will regularly replenish the supply of massive stars in these regions.
This naturally leads to the expectation of frequent core-collapse supernovae in such environments.
As an example, \cite{Zubovas13} show that, per $10^6~\Msun$ of stellar mass formed in the Galactic Centre, approximately one supernova per $10^{4}~\mathrm{yr}$ is expected for the past $10^8~\mathrm{yr}$.

Only one supernova remnant (SNR) candidate has been identified close to the SMBH sphere of influence (SOI): an elongated shell known as Sgr A East, at the end of its adiabatic phase.
It has an estimated age of about $10^4~\mathrm{yr}$ and appears to be engulfing Sgr A* with a mean radius of approximately $5~\mathrm{pc}$ \citep{Maeda02, Herrnstein05, Lee06, Tsuboi09}.
In addition, there are a couple of observations that indirectly point towards supernovae in the SOI.
The first is CXOGC J174545.5--285829 (`The Cannonball'), suspected to be a runaway neutron star associated with the same supernova explosion as Sgr A East \citep{Park05, Nynka13, Zhao13}.
The second is the recently discovered magnetar SGR~J1745--2900, estimated to be within $2~\mathrm{pc}$ of Sgr A* \citep[][]{Degenaar13, Kennea13, Rea13}.

Any supernova exploding in the SOI of a quiescent SMBH will expand into a gaseous environment constituted mainly by the SMBH accretion flow, whose gas is supplied by the winds from massive stars.
The density distribution within the flow is therefore set by both the number and distribution of young stars and the hydrodynamical properties of a radiatively inefficient accretion regime.
This interplay gives an overall density distribution that is a broken power law, for which the break occurs where the number density of stellar wind sources drops off.
For the Galactic Centre, this corresponds to $\sim 0.4~\mathrm{pc}$ \citep[for example,][]{Quataert04}.

In such environments, we expect SNRs to evolve differently from those in the typically flat interstellar medium, away from the SMBH.
The density gradients have the potential to distort SNRs and decelerate them significantly.
Once the expansion velocity falls below the SMBH velocity field, the remnant will be tidally sheared and eventually torn apart.
This can substantially shorten an SNR lifetime compared to that in a constant-density interstellar environment.
In turn, this can reduce the expected number of observed SNRs in galactic nuclei.  

Since quiescent accretion flows are fed by stellar winds, which can be also partially recycled to form new stars together with the gas released by supernova explosions, the scenario we consider is of a self-regulating environment, where young stars and gas (or, in other words, star formation and accretion on to the SMBH) are intimately related.
This holds until a violent event---for example, a merger---drives abundant stars and gas from larger scales to the galactic nucleus.
Observations and modelling of our Galactic Centre support this picture.
In particular, winds from massive stars are sufficient to account for the observed accretion luminosity and external gas feeding is not required \citep[e.g.][]{Quataert04,Cuadra06} or observed.
Furthermore, there is strong evidence for the recent star formation occurring in situ \citep{Paumard06}.

In this paper, we determine the morphology and X-ray lifetimes of SNRs, which, in turn,  can be used to constrain the environment of SMBHs.
We develop a numerical method to trace SNR evolution and determine their X-ray lifetime.
The influence of the SMBH on SNRs will be considered first indirectly, through its influence on the gaseous environment, and then directly, through its tidal shear of the ejecta. 

The paper is organized as follows.
Section~\ref{sec:quiescent_environments} introduces the gaseous environments found around quiescent SMBHs.
Section~\ref{sec:method_analytic} uses analytic methods to qualitatively trace SNR evolution.
Section~\ref{sec:method_numerical} describes our numerical method, which allows us to follow the evolution of an SNR in an arbitrary axially symmetric gas distribution.
We then specialize it to a quiescent SMBH environment.
Section~\ref{sec:method_nuclei} outlines the galactic models used for the environments of the supernova simulations.
Section~\ref{sec:results} presents our results for SNR shapes and lifetimes.
Our concluding remarks are found in Section~\ref{sec:conclusion}.

\section{Gaseous environments of quiescent galactic nuclei} \label{sec:quiescent_environments}
In this section, we outline the expected gas distributions near the SMBH in quiescent galactic nuclei.
These gas distributions will be used as the environment for the SNR model exposited in Sections \ref{sec:method_analytic} and \ref{sec:method_numerical}.
We will then proceed to scale the general environment discussed here for the Galactic Centre to other SMBHs in Section~\ref{sec:method_nuclei}.

Quiescent SMBHs are surrounded by RIAFs, which are the environments in which the SNR will evolve.
RIAFs are relatively thick, for which the scale-height, $H$, is comparable to the radial distance, $R$, from the SMBH ($H/R \approx 1$).
The mechanisms of energy transport within the flow vary depending on the model, and these variations affect the power-law gradient in density near the SMBH.
Advection-dominated accretion flow (ADAF) models assume that much of the energy is contained in the ionic component of a two-temperature plasma.
As the ions are much less efficient radiators than electrons, energy is advected into the SMBH by the ions before it can be lost via radiation \citep{Narayan95a, Narayan95b}.
Additionally, convection-dominated accretion flow (CDAF) models rely on the transport of energy outward via convective motions in the gas \citep{Quataert00, Ball01}.
Finally, the adiabatic inflow--outflow model \citep[ADIOS;][]{Blandford99, Blandford04, Begelman12} accounts for winds from the flow that expel hot gas before it is accreted.

For the region near the SMBH, predicted exponents, $\omega_\mathrm{in}$, of the power law in gas density, $\rho$, lie in the range of $\omega_\mathrm{in} = 1/2$ to $3/2$.
The lower and upper limits of $\omega_\mathrm{in}$ are derived from the predictions of the CDAF/ADIOS and ADAF models, respectively.
A drop-off in stellar number density at a radius $R = R_\mathrm{b}$ from the SMBH would cause a break in the mass density, $\rho$, at the same radius, since it is the winds from these stars that feed the accretion flow.

The best example of a RIAF is that surrounding Sgr A*.
It has been extensively studied theoretically and observationally and will constitute our prototype.
A density distribution from the one-dimensional analytic model of wind sources has approximately a broken power-law shape with $\omega_\mathrm{in} = 1$ inside the density break and $\omega_\mathrm{out} = 3$ outside \citep{Quataert04}.
Simulations of stellar wind accretion show comparable density profiles \citep{Cuadra06}.
Furthermore, the value of $\omega_\mathrm{in} = 1$ is consistent with GRMHD accretion simulations \citep[for example,][]{McKinney12}.
Recent observations using long integrations in X-ray suggest that a gradient of $\omega_\mathrm{in} \approx 1/2$ may provide a better fit to the inner accretion flow of Sgr A* \citep{Wang13}.

We can therefore, generally describe the ambient medium of a quiescent SOI with a broken power law for the density of the form:
\begin{equation}
\rho(R) =
\begin{dcases*}
\rho_0 \left( \frac{R}{R_0} \right)^{-\omega_\mathrm{in}} & $R \leq R_\mathrm{b}$ \\
\rho_\mathrm{b} \left( \frac{R}{R_\mathrm{b}} \right)^{-\omega_\mathrm{out}} & $R > R_\mathrm{b}$,
\end{dcases*}
\end{equation}
for $\omega_\mathrm{in} \in \{1/2, \, 1, \, 3/2\}$, $\omega_\mathrm{out} = 3$, using a reference point for the density at $R = R_0$ away from the SMBH.

The strongest observational constraint on the density around Sgr A* is given by \textit{Chandra} X-ray measurements at the scale of the Bondi radius ($R_0 \approx 0.04~\mathrm{pc}$) of $n_0 \approx 130~\mathrm{cm}^{-3}$ \citep[$\rho_0 \approx 2.2 \times 10^{-22}~\mathrm{g \, cm}^{-3}$;][]{Baganoff03}.
The accretion rate closer to the SMBH can be further constrained by Faraday rotation measurements, though the relative error is large \citep{Marrone07}.
Indeed, we find that fixing the density at $0.04~\mathrm{pc}$ and varying $\omega_\mathrm{in}$ between $1/2$ and $3/2$ produces a range of densities at small radii that fall within the uncertainty in the density inferred from Faraday rotation.
The radius for the break in stellar number density and gas density in the Milky Way is taken to be $R_\mathrm{b} = 0.4~\mathrm{pc}$.

\section{Evolution of remnants around quiescent black holes: analytic foundations} \label{sec:method_analytic}
Here, we outline the physics describing the early stages of SNR evolution that are of interest in this work.
The theory described in this section will be used as the foundation of a general numerical method to solve the problem, outlined in Section~\ref{sec:method_numerical}.
At this point, we do not {directly} take into account the gravitational force of the SMBH, but instead just the gaseous environment.
The gravity of the SMBH can be ignored when the expansion velocity of the SNR is much larger than the Keplerian velocity around the SMBH.
For example, around Sgr A*, at a velocity of $10^4~\mathrm{km \, s}^{-1}$ gravity can be ignored for radii larger than $\sim 10^{-4}~\mathrm{pc}$.
The gravitational field of the SMBH will be accounted for later, when we consider tidal effects on the expanding remnant, which are important only once the remnant has slowed down significantly.

A supernova explosion drives a strong shock into the surrounding gas at approximately the radial velocity of the ejected debris.
Typically, it is assumed that a significant amount of the ejecta is contained within a shell just behind the shock front \citep[for example,][]{Koo90}.
As it expands, the shock sweeps up further mass from the surrounding medium.
By momentum conservation, the combined mass of the fraction of ejecta behind the shock front ($M_\mathrm{ej}$) plus the swept-up gas ($M_\mathrm{s}$) must decelerate.
The deceleration is considered to be appreciable when the swept-up mass becomes comparable to that of the debris, and therefore this ejecta-dominated phase holds for $M_\mathrm{s} \ll M_\mathrm{ej}$.

The subsequent adiabatic expansion of the shock front is modelled with the assumption that losses of energy internal to the remnant are negligible.
For this decelerating regime, the Rankine--Hugoniot strong-shock jump conditions can yield exact similarity (length scale-independent) solutions for the kinematics of the shock front.
The evolution is determined by its energy, $E$, and the ambient density, $\rho$ \citep{McKee95}.
In all of this work, we use a canonical value of $10^{51}~\mathrm{erg}$ for the explosion energy.
In a uniform ambient medium, the adiabatic stage is classically modelled using the spherically symmetric Sedov--Taylor solution \citep{Taylor50, Sedov59}.
This has self-similar forms for the spherical radius and speed of the SNR of $R' \propto \left(E/\rho\right)^{1/5} t^{2/5}$ and $v \propto \left(E/\rho\right)^{1/5} t^{-3/5}$, respectively, where $R'$ is measured from the explosion site.

Following the initial work by Sedov and Taylor, \cite{Kompaneets60} developed a non-linear equation from the jump conditions that allows self-similar solutions for the shock front evolution in certain density stratifications.
The original work by Kompaneets considered an atmosphere with exponential stratification, but many other solutions have since been obtained (see the review by \citealt{BS93}, as well as \citealt{Bannikova12} and the references therein).
Of particular relevance to the gas distributions in galactic nuclei, \cite{K92}---hereafter, K92---showed that, with a specific coordinate transformation, a circular solution to the Kompaneets equation can be obtained for explosions offset from the origin of a power-law density profile, $R^{-\omega}$ (for $\omega \neq 2$).

The early ejecta-dominated and late adiabatic stages are well characterized by the purely analytic solutions for each stage.
In between, the solution asymptotically transitions between these two limits \citep[this is known as `intermediate-asymptotic' behaviour;][]{Truelove99}.\footnote{For an illustration of this transition, see fig.~2 of \cite{Truelove99}.}
The late evolution of the remnant, the radiative stage, occurs when the temperature behind the shock drops to the point at which there is an appreciable number of bound electrons.
Consequently, line cooling becomes effective, the radiative loss of energy is no longer negligible, and the speed of the shock will drop at a faster rate.
For SNRs in a constant density of $n \approx 1~\mathrm{cm}^{-3}$, the radiative phase begins at approximately $3 \times 10^4~\mathrm{yr}$ \citep{Blondin98}.
We do not model the remnant during this phase, but we will estimate the onset of the transition to the radiative stage.

In the present work, we model SNRs over the first two (ejecta-dominated and adiabatically expanding) stages of evolution in a range of galactic nuclear environments.
The evolution begins with a spherically expanding shock, and therefore we do not consider any intrinsic asymmetries in the supernova explosion itself.
Collectively, any possible intrinsic asymmetries in SNRs are not expected to be in a preferential direction, and so they should not bias the generalized results presented here.

\begin{figure*}
\begin{center}
\includegraphics[width=1.2\columnwidth]{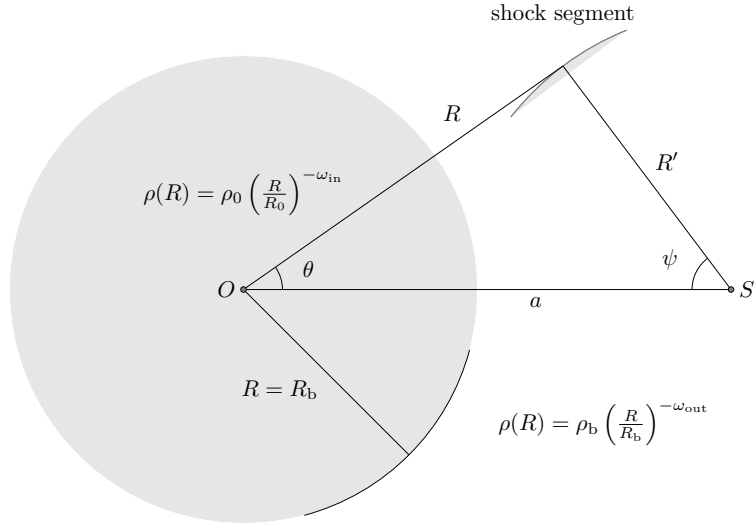}
\end{center}
\caption{Basic geometry of the problem.
The supernova occurs at a point $S$, a distance $R = a$ away from the SMBH, which is located at the origin, $O$.
The shock front extends to distances measured radially from the explosion point $S$ by the coordinate $R'$.
The angle made by a point on the shock, measured from the $\theta = 0$ axis about the explosion point, is denoted $\psi$.
Each point on the shock has an initial angle $\psi (t \rightarrow 0) \equiv \psi_0$.
The entire density distribution $\rho(R)$ can be characterized by: the choice of the inner gradient $\omega_\mathrm{in}$ (defining the density within the shaded circle), the outer gradient $\omega_\mathrm{out}$, the reference density $\rho_0$ (at a reference radius $R_0$), and a break at $R_\mathrm{b}$ between the gradients $\omega_\mathrm{in}$ and $\omega_\mathrm{out}$.}
\label{fig:geometry}
\end{figure*}

The overall geometry of this analysis is laid out in Fig.~\ref{fig:geometry}, which indicates the main coordinates, distance scales and density distributions.
The explosion point is at a distance $R = a$, measured from the SMBH (the origin of our coordinate system).
The shock front extends to radial distances $R'$, measured from the explosion point.
Each point along the shock is at an angle $\psi$, measured from the axis of symmetry about the explosion point.
The initial angle made with the axis of symmetry of each point on the shock, at $t \rightarrow 0$, is denoted $\psi_0$.

\subsection{End of the ejecta-dominated stage} \label{sec:method_deceleration_lengths}
In order to estimate where the shock front kinematics appreciably deviate from the ejecta-dominated solution, we integrate the background density field along spherical volume elements swept out by the expanding remnant.
This provides an estimate of the mass swept up from the environment, $M_\mathrm{s}$.
The ejecta-dominated solution is taken to end when $M_\mathrm{s}$ is equal to some specified portion of the ejecta mass, $M_\mathrm{ej}$.
We use a canonical value of $1 \Msun$ for this fraction of ejecta mass.
The distance from the explosion point (along the coordinate $R'$) at which this occurs is denoted the `deceleration length', $L$, here (it also known as the `Sedov Length' in the standard treatment of SNRs in a uniform $\rho$).

Since our density profiles are not uniform, different directions of expanding ejecta will sweep up mass at different rates.
In general, we must consider a solution for $L$ that depends on $\psi_0$, the initial angle of each surface element of the shock with respect to the axis of symmetry (see Fig.~\ref{fig:geometry}).
We therefore determine the value of $L(\psi_0)$ corresponding to small surface elements of the shock front.
When the explosion occurs close to the SMBH, the solution is expected to converge to that of an integral over a sphere, due to the spherical symmetry of the background density.\footnote{The three-dimensional volume integrals (of an offset sphere) over a singular density converge for the shallow power laws used here: $1/2 \leq \omega_\mathrm{in} \leq 3/2$ for $\rho \propto R^{-\omega_\mathrm{in}}$.}
Therefore, as a reference, we also find the radius $L$ of the sphere whose volume encloses $M_\mathrm{s} \approx M_\mathrm{ej}$.

The explosion occurs at a distance $R = a$ from the origin.
For a single power-law stratification, we use the explosion point for the reference density, $\rho_0 = \rho(a) \equiv \rho_a$, such that
\begin{equation} \label{eq:density_power_law}
\rho(R) = \rho_a \left(\frac{R}{a}\right)^{-\omega} .
\end{equation}
We consider a small surface element of the SNR at an angle $\psi_0$ over an infinitesimal solid angle.
In a single power-law stratification with the form of equation (\ref{eq:density_power_law}), the length $L\left(\psi_\mathrm{0}\right)$ can be estimated from the mass integrated through $R'$ at a given angle $\psi_0$:
\begin{equation} \label{eq:solid_angle_integral}
\rho_a a^\omega \int_{0}^{L(\psi_0)} R^{-\omega} R'^{\, 2} \, \ud R' = M_\mathrm{ej}.
\end{equation}
Note that we are integrating over the coordinate $R'$ that extends radially from the explosion point, but that the density varies radially with the coordinate $R$ as measured from the SMBH.
For integrals over a broken power-law density, the density break adds complications to the integrals analogous to equation (\ref{eq:solid_angle_integral}).
The solutions are discussed further in Appendix \ref{ap:integrals}.

These methods for estimating the deceleration length provide a means for testing the level of asymmetry and distance scales in the ejecta-dominated stage of evolution, and will be further discussed in Section~\ref{sec:results_deceleration_lengths}, where we show results.

\subsection{Deceleration in the adiabatic stage} \label{sec:method_kompaneets}
We use the \cite{Kompaneets60} approximation alongside the coordinate transformation identified by K92 to follow the adiabatic deceleration of the shock front in a single power-law density profile.
The assumptions and main equations of this prescription will also be used in our full numerical treatment for arbitrary density profiles (Section~\ref{sec:method_numerical}).
We shall give here the analytic solutions for $\omega = 1$ and $3$.
These solutions will be used to validate our numerical treatment (Section~\ref{sec:method_numerical}).
They also give an indication of the shock behaviour in a broken power-law density profile, when it expands fully interior or fully exterior to the density break. 

The Kompaneets approximation involves setting the post-shock\footnote{For thermodynamic variables, we use primes ($'$) to indicate the post-shock values (the values behind the shock front).} pressure, $P'$, to be uniform throughout the shock volume and equal to (some fraction, $\lambda$, of) the mean interior energy density.
For an arbitrary volume $V$,
\begin{equation}
P' = \frac{\left(\gamma - 1\right) \lambda E}{V} ,
\end{equation}
wherein the Kompaneets approximation proper is to take $\lambda$ to be constant.
The ratio of specific heats is taken to be $\gamma = 5/3$ both internal and external to the shock.

Two additional assumptions in the treatment are that the directions of the local velocity vectors along the shock front are normal to the shock front, and that the magnitude of the velocity is determined by taking the post-shock pressure to be equal to that of the ram pressure of the environment ($\rho v_\mathrm{s}^2$, where $\rho$ is the density of the unshocked gas) at that point (K92):
\begin{equation} \label{eq:velocity}
v_\mathrm{s}(R,t) = \sqrt{\frac{\left(\gamma^2 - 1\right) \lambda E}{2\rho(R) \, V(t)}} .
\end{equation}

Following the coordinate transformation of K92, the `time' is parametrized by $y$ (which actually has a dimension of length) via
\begin{equation}
\label{eq:dydt}
\ud y = \sqrt{\frac{\left(\gamma^2 - 1\right)\lambda E}{2 \rho_0 V(y)}} \, \ud t ,
\end{equation}
as well as the dimensionless parameter $x = \abs{2 - \omega} y / \left( 2a \right) \equiv y / y_\mathrm{c}$.
The parameter $x$ is, therefore, equal to $y$ scaled with respect to a critical value $y_\mathrm{c}$, which is when the shock either reaches the origin ($\omega = 1$) or `blows out' to infinity ($\omega = 3$).
Therefore, $x$ (like $y$) can be considered to represent the `time' in this transformation.
The constant $\lambda \approx 1$ is given by the difference in pressure behind the shock front relative to the average pressure internal to the remnant, and in a power-law profile is
\citep{Shapiro79}
\begin{equation} \label{eq:lambda}
\lambda = \frac{\left(17 - 4\omega\right)/9}{1 - \left(9 - 2\omega\right)^{-\left(17 - 4\omega\right)/12 - 3\omega}} .
\end{equation}

In an ambient density with a single power-law form of equation (\ref{eq:density_power_law}), the K92 transformation gives a self-similar solution to the Kompaneets equation (see equations\footnote{Note that there are two sign errors in the exponents of equation 11 in K92.} 10 and 11 of K92) for an explosion at $R = a$ (see Fig.~\ref{fig:geometry}):
\begin{equation} \label{eq:Psi}
\left(\frac{R}{a}\right)^{2\alpha} - 2\left(\frac{R}{a}\right)^{\alpha} \cos{(\alpha\theta)} - x^2 + 1 = 0 ,
\end{equation}
for the polar coordinates $R$ and $\theta$, where $\alpha \equiv \left(2 - \omega\right)/2$.
This can be identified as a circular solution for a given $x$ in the two variables $\left(R/a\right)^\alpha$ and $\alpha \theta$.
Analytic solutions for the volume, time and velocity in $\omega = 1$ and $3$ densities are presented in Appendix \ref{ap:kompaneets_solutions}.

The equations describing the shock front can alternatively be parametrized by $\psi_0$, the initial angle of a point on the shock with respect to the axis of symmetry.
The subsequent equations of motion for a given $\psi_0$ describe the paths of flowlines in the shock in terms of the polar coordinates measured from the SMBH (K92):
\begin{align}
R      &= a \left(1 + 2 x \cos{\psi_0} + x^2\right)^{1/\left(2 \alpha \right)} , \label{eq:psiR} \\
\theta &= \frac{1}{\abs{\alpha}} \arctan{\left(\frac{x \sin{\psi_0}}{1 + x \cos{\psi_0}}\right)}. \label{eq:psiT}
\end{align}
This flowline-based treatment is a useful context for the numerical approach to the shock evolution presented in Section~\ref{sec:method_numerical_scheme}, and these equations will be used to compare with the numerical results.

\subsection{Intermediate-asymptotic transition} \label{sec:method_density}
The Kompaneets solution for the velocity diverges for $x \rightarrow 0$, given that the volume $V(x) \rightarrow 0$.
In this limit, the energy density and, therefore, also the velocity, tend to infinity.
The solution is, however, not intended to describe the initial evolutionary stage of the remnant.
In order for the numerical treatment to correctly follow the SNR evolution, we must account for the initial coasting stage.
A full analytic joining of the intermediate-asymptotic solutions between the ejecta-dominated and adiabatic stages is complex, even for an $\omega = 0$ ambient medium \citep[see, for example,][]{Truelove99}.

As a model for this intermediate behaviour, we employ an effective density (mass) term to the solution that gives a transition between the expected solutions.
The density of the medium is modified to:
\begin{equation} \label{eq:rhoeff}
\rho_\mathrm{eff} \equiv \rho(R) + \frac{M_\mathrm{ej}}{V} ,
\end{equation}
where the additional effective term counters the divergent behaviour of the velocity at small volumes.
This has the desired property that when the volume is large $\rho \rightarrow \rho(R)$ in the standard Kompaneets approximation, while at small volumes the second term dominates to provide the initial coasting phase of the remnant.
With this effective mass term, the SNR leaves the ejecta-dominated phase around the point at which the mass swept up from the environment is comparable to the initial mass behind the shock.

\subsection{Transition to the radiative stage} \label{sec:method_radiative}
As the shock slows, the late evolution of a typical SNR is marked by an increase in radiative losses.
Although we will not model this stage, we intend to check the time-scales over which SNRs will reach this stage in quiescent nuclei (if they survive sufficiently long).

Typically, cooling functions show a marked increase in thermal radiation once the gas temperature drops to $\sim 10^6~\mathrm{K}$ \citep[for example,][]{Schure09}.
This occurs due to the formation of a sufficient number of electrons bound to ions to allow for effective line cooling.
Once regions of gas behind the shock drop to this temperature, the deceleration of the SNR becomes more pronounced.
By calculating the temperature behind the shock we can determine the time at which parts of the remnant begin to cool more effectively.

It is possible to determine the temperature of the shocked gas via the ideal gas law,
\begin{equation}
P' = \frac{k_\mathrm{B} \rho' T'}{m_u \mu}
\end{equation}
(where, again, we denote post-shock values with primes, $k_\mathrm{B}$ is Boltzmann's constant, $m_u$ is the atomic mass unit and $\mu$ is the mean molecular mass), as well as the jump conditions for the (post-shock) density and pressure,
\begin{equation}
\rho' = \rho \, \frac{\gamma + 1}{\gamma - 1} \quad \mathrm{and} \quad P' = \frac{2 \rho v_\mathrm{s}^2}{\gamma + 1} .
\end{equation}
For $\gamma = 5/3$,
\begin{equation}
T' = \frac{2 \left(\gamma - 1\right) m_u \mu}{\left(\gamma + 1\right)^2 k_\mathrm{B}} v_\mathrm{s}^2 = \frac{3 m_u \mu}{16 k_\mathrm{B}} v_\mathrm{s}^2 ,
\end{equation}
and the post-shock temperature is found to be $T' \approx 10^6~\mathrm{K}$ for $v_\mathrm{s} \approx 300~\mathrm{km \, s}^{-1}$.
Therefore, if we monitor each point along the shock for the time at which the velocity drops below this value, we may estimate the time at which radiative processes become significant.

As the cooling function is also dependent on $\rho$, for a given temperature the rate of cooling is also expected to be amplified in regions of post-shock material with higher density.
However, by the time that SNRs are radiative, they have survived the expansion past the SMBH and entered into the more uniform density beyond the SOI, such that the ambient density is similar across all points of the shock.
At this stage, the SNRs are reasonably symmetric around the SMBH and the velocity is similar across all of the shock front, so that most of the SNR reaches the radiative stage at similar times.
If the SNR survives expansion past the SMBH, this late evolution is largely uninfluenced by the details of any early interactions near the SMBH.\footnote{For a detailed consideration of the radiative transition in power-law media, see \cite{Petruk06}.}.

\section{Evolution of remnants around quiescent black holes: numerical treatment} \label{sec:method_numerical}
Purely analytic solutions for the shock front evolution via the Kompaneets equation are not feasible for many density configurations.
Therefore, we developed a numerical method that solves for the evolution of a shock front using the physical assumptions of the Kompaneets approximation described in Section~\ref{sec:method_analytic}.

The primary assumptions that must be encompassed by the method culminate in constraints on the velocity.
Namely, the direction of the velocity of any point must be perpendicular to the shock front, and the magnitude of the velocity must be determined by the energy density behind the shock and local ambient density as prescribed in equation (\ref{eq:velocity}).

\subsection{General prescription} \label{sec:method_numerical_scheme}
The numerical treatment follows an approach by which the shock is described by the evolution of flowlines through the background gas.
The flowlines are the paths followed by tracer `particles' (points) distributed along the shock front, analogous to the analytic treatment with the $\psi_0$ parameter of equations (\ref{eq:psiR}) and (\ref{eq:psiT}).

\begin{figure}
\begin{center}
\includegraphics[width=\columnwidth]{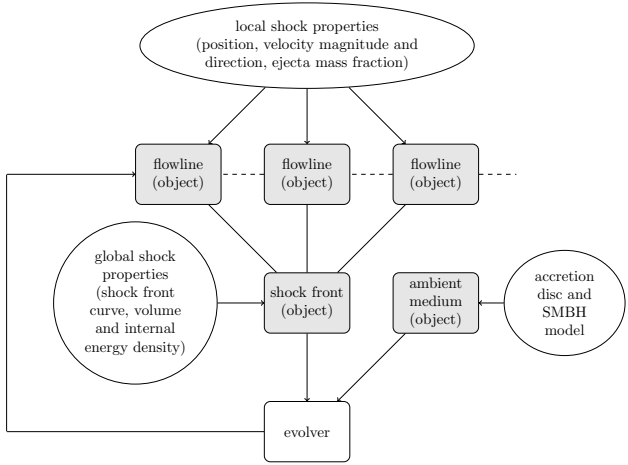}
\end{center}
\caption{The basic numerical scheme, as described in Section~\ref{sec:method_numerical_scheme}.
Shaded boxes show the basic types of objects in the numerical construction.
The choice of the number of flowlines determines the resolution, and only a few are shown here schematically.
The flowlines track local physical properties of the shock (velocity and ejecta mass fraction).
Collectively, they define the location of the shock front, with global physical properties such as its energy density, determined by the volume.
Along with the environment (most importantly, the background mass density), the global shock properties determine the evolution of all the individual flowlines in the subsequent time step.}
\label{fig:numerical_structure}
\end{figure}

Fig.~\ref{fig:numerical_structure} shows a schematic of the approach.
The initial (spherical) state of the shock is broken down into flowlines characterized by their angle $\psi_0$.
During the evolution, the number of flowlines is dynamic.
To keep a reasonable resolution of the shock front, new flowlines can be inserted, with mean properties of adjacent flowlines, if the distance between two points on the shock is over a defined threshold.
For our simulations, a threshold of the order of $0.05~\mathrm{pc}$ has proven sufficient to describe a smooth shock front evolution on Milky Way-like scales.
Flowlines may also be deleted in regions where parts of the shock front are colliding.
The background gas prescribes the evolution of the shock, but the behaviour of the post-shock gas is not tracked, and thus the background gas can be treated as being independent of the shock.

The kinematics of the shock front are determined by the velocity vectors at each flowline.
To determine the magnitude of the velocity, we use the jump conditions across the shock, and for those we need the energy density within the shocked volume (which is assumed to be a constant within this volume) as well as the local mass density in the environment (see equations \ref{eq:velocity} and \ref{eq:rhoeff}.\footnote{For simplicity, the ratio $\lambda \approx 1$ of the post-shock pressure to mean interior energy density is set to be exactly unity in equation (\ref{eq:velocity}). As the shock velocity is proportional to $\sqrt{\lambda}$, the effect of this is small compared to other limitations inherent in the Kompaneets approximation discussed in Section~\ref{sec:method_limitations}.}
This stage of evolution is adiabatic, and so given an initial explosion energy we therefore calculate the energy density using the instantaneous volume enveloped by the shock front.

In an axisymmetric arrangement of gas density and explosion point, calculation of the volume is simplified by the geometrical symmetry; it is determined by a solid of rotation of the area of a two-dimensional slice about this axis.
Any arbitrary ordered set of points $\left(x_i, y_i \right)$ can specify the location of the shock front.
Given these two-dimensional coordinates, the volume, by the second theorem of Pappus, is equal to the product of the area of the non-intersecting polygon defined by these coordinates and the distance travelled by its centroid under rotation about the symmetry axis \citep{Kern48}.
Using the fact that the components of the centroid, $C = \left(C_x, \,  C_y \right)$, of a polygon are given by
\begin{equation}
C_\xi = \frac{1}{6A} \sum_{i=1}^{n-1}{\left(\xi_i + \xi_{i+1}\right) \left(x_i y_{i+1} - x_{i+1} y_i\right)}
\end{equation}
for $\xi \in \left\{ x, y \right\}$, the volume of the SNR can be determined from
\begin{equation} \label{eq:volrotation}
V = \frac{\uppi}{3} \sum_{i=1}^{n-1}{\left(y_i + y_{i+1}\right) \left(x_i y_{i+1} - x_{i+1} y_i\right)} .
\end{equation}

With the magnitude of the velocity known, each point on the shock evolves by determining the unit vector for the velocity that is perpendicular to its neighbouring points.
The position is then linearly translated over a small time step using the velocity vector.

\begin{figure}
\begin{center}
\includegraphics[width=\columnwidth]{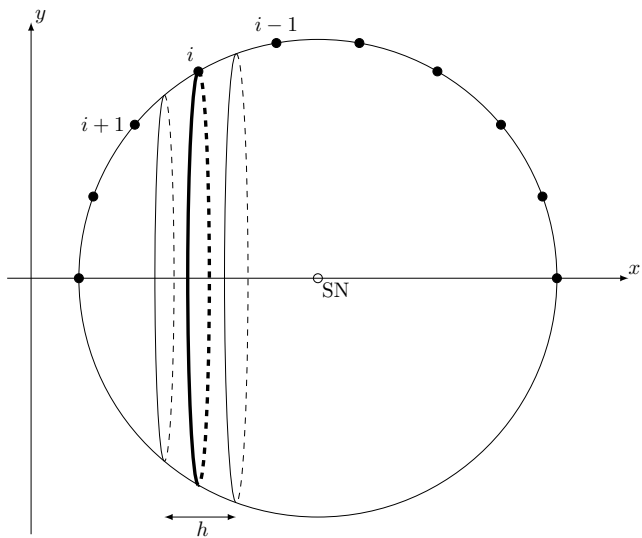}
\end{center}
\caption{A schematic of the initial, spherical state of the SNR, with initial positions for the flowlines (filled circles) around the point of explosion (open circle labelled `SN').
Flowlines in the positive-$y$ portion of the $x$--$y$ plane define a three-dimensional shock front by rotation about the axis of symmetry.
Rotating each flowline about this axis produces a ring (shown for the $i^{\mathrm{th}}$ flowline as a thick line).
The midpoints between the $i^{\mathrm{th}}$ flowline and its neighbours define the limits of the zone of the sphere assigned to that flowline (thin lines).
For a sphere of radius $R'$, the area of this zone is proportional to the height of the zone, $h$, since its surface area is $2 \uppi R' h$.
The fraction of total ejecta mass assigned to the flowline is then the ratio of this area to the total area of the sphere.}
\label{fig:sphere_flowlines}
\end{figure}

We assign an ejecta mass element to each flowline.
Fig.~\ref{fig:sphere_flowlines} shows a schematic of this implementation.
Due to rotational symmetry, each point on the shock at $t \rightarrow 0$ represents a ring segment of the SNR in three dimensions.
We assign a thickness to each of these ring segments based on the spacing between the flowlines in the initial spherical state.
The fraction of ejecta mass represented by the flowline is then the ratio of the area of this zone of the sphere to the total surface area of the sphere.

Due to asymmetry in the background density and the presence of a strong density contrast near the origin, segments of the shock may collide with one another.
This shock front self-interaction can lead to collisions in which kinetic energy is converted into internal energy.
Since it cannot be easily radiated away, we expect a transient acceleration outward of the heated gas, after which the fluid will return to the dynamics imposed by the global expansion.
The numerical treatment of these self-interactions is outlined in Appendix \ref{ap:self_interactions}.
This treatment results in the deletion of some flowlines, accounting for the modification of the flow in this region.

\subsection{Comparison with analytic solutions for single power-law profiles} \label{sec:method_compare_analytic_numerical}
\begin{figure*}
\begin{center}
\includegraphics[width=1.7\columnwidth]{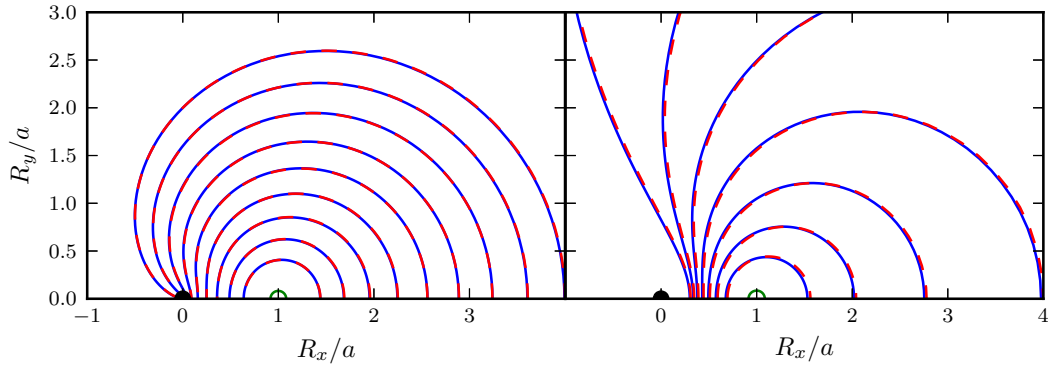}
\end{center}
\caption{Numerical results (blue) compared with analytic solutions (red, dashed) for the locations of the shock front in density profiles with $\omega =1$ (left-hand panel) and $\omega = 3$ (right-hand panel).
The initial (spherical) state for the numerical solutions is shown in green.
Units are given as ratios of distance $\left(R_x, R_y\right)$ to explosion distance $a$ from the density singularity at $\left(0, 0\right)$ (the SMBH in our model).
The results can be written in parametric form in terms of $x$, which increases with time $t$ up to a critical value of $x = 1$; see Section~\ref{sec:method_kompaneets} as well as the expressions for $t(x)$ in Appendix \ref{ap:kompaneets_solutions}.
The solutions for $\omega = 1$ are found up to $x = 1$, while for $\omega = 3$ they are given up to $x = 0.8$ due to the divergence of solutions as $x \rightarrow 1$ in this latter case.
The trailing part of the shock (the part directly towards the SMBH) for $\omega = 1$ solution reaches $R_\mathrm{x}/a = 0$ at $x = 1$, while leading point (directly away from the SMBH) reaches $R_\mathrm{x}/a = 4$.
The trailing part of the $\omega = 3$ solution asymptotically approaches a distance of $R_\mathrm{x}/a = 1/4$ as $x \rightarrow 1$, while in the same limit the leading part of the shock follows $R_\mathrm{x}/a \rightarrow \infty$.}
\label{fig:compare_analytic_numerical}
\end{figure*}

Fig.~\ref{fig:compare_analytic_numerical} shows the morphology of an SNR, running into a circum-SMBH environment with density power-law gradient $\omega = 1$ (left-hand panel) and $\omega = 3$ (right-hand panel).
There, we compare the analytic prescription described in Section~\ref{sec:method_kompaneets} and Appendix \ref{ap:kompaneets_solutions} with our numerical method.
The numerical solutions are found to match the analytic form very well.
There is very slight deviation between the two methods, more noticeably in the $\omega = 3$ case, which is due to the fact that the numerical method requires a small spherical initial step.
The analytic solution is closer to a sphere at small times in the $\omega = 1$ solution so there is almost no discernible discrepancy.

\begin{figure*}
\begin{center}
\includegraphics[width=2.1\columnwidth]{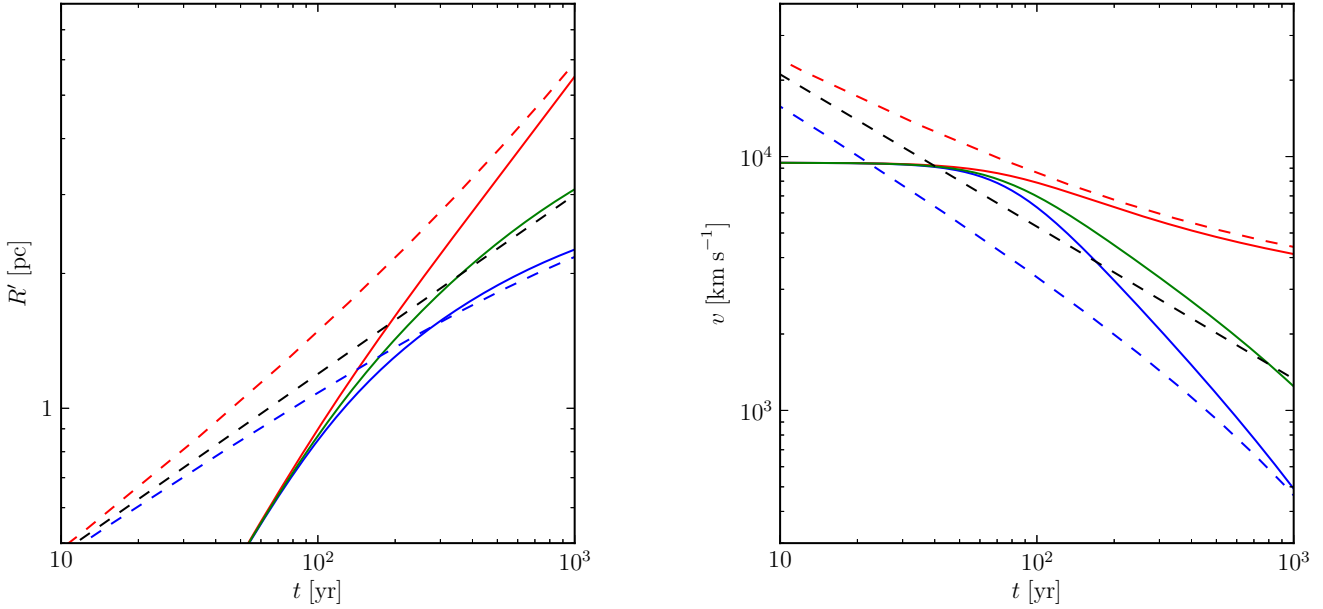}
\caption{Example of radius and velocity evolution for a remnant in a broken-power-law medium (solid lines).
The explosion occurs at $3~\mathrm{pc}$ (near the density break between $\omega_\mathrm{in} =1$ and $\omega_\mathrm{out} = 3$) around a $5 \times 10^8 \Msun$ SMBH using the scaling described in Section~\ref{sec:method_nuclei}.
Solid lines are plotted from snapshots of the evolution of the remnant in the numerical treatment, where blue (lowermost) curves are for the trailing flowline (towards the SMBH) and red (uppermost) curves are for the leading one (away from the SMBH).
The green curve shows the behaviour of a flowline in the numerical treatment that emerges at $90\degree$ from the $\theta = 0$ axis ($\psi_0 = \uppi/2$).
Some analytic results (with arbitrary scaling) are given as dashed lines for comparison.
Black is the form of the Sedov-Taylor (uniform medium, $\omega=0$) solutions.
The red dashed curve shows the solution for the point on the shock travelling directly away from the SMBH for a shock in a purely $\omega = 3$ medium.
The blue dashed curve shows the solution for the point on the shock travelling directly towards the SMBH in a purely $\omega = 1$ medium.}
\label{fig:v_r_compare}
\end{center}
\end{figure*}

Fig.~\ref{fig:v_r_compare} shows a comparison between our numerical (solid lines) and the analytic (dashed lines, arbitrary scaling) results in a broken power-law medium with $\omega_\mathrm{in} = 1$ and $\omega_\mathrm{out} = 3$.
This figure shows the distance and velocity evolution of selected sample points on the shock front.
We follow the portion of the shock that propagates towards the SMBH (blue lines), away from the SMBH (red lines) and at an initial angle of $\psi_0 = \uppi/2$ (green line).
The numerical radius and velocity are seen to transition from the coasting (radius $R' \propto t$, velocity $v = \mathrm{const.}$) phase to forms similar to those seen in the pure Kompaneets solutions.

As expected, the evolution of the trailing part of the shock, as it gets closer to the SMBH, approaches the analytic solution for a pure $\omega = 1$ medium.
Likewise, the leading part of the shock asymptotes to the pure $\omega = 3$ analytic solution, as it expands away from the SMBH.
The green line shows how the evolution of a flowline that emerges at $90\degree$ from the $\theta = 0$ axis has, instead, an intermediate behaviour, which is influenced by the overall broken power-law density.
The figure also shows for reference the Sedov--Taylor solution \citep[black dashed line,][]{Taylor50, Sedov59} for an explosion in a uniform ambient medium ($R' \propto t^{2/5}$ and $v \propto t^{-3/5}$).

\subsection{Caveats and limitations of the model} \label{sec:method_limitations}
For more complex background density configurations, such as one with many large density contrasts that trigger self-interactions and turbulence, one may consider a treatment of self-interacting shocks that is more in-depth and sophisticated than that presented in Appendix \ref{ap:self_interactions}.
The increase in velocity of any small self-intersecting region is expected to be a brief transient phenomenon; therefore, we do not presently apply any boost in velocity when merging flowlines, instead only accounting for the net direction of the flow that results from two colliding parts of the shock.
The reason is that, in all our simulations, the portion of the shock front which undertakes self-interaction is limited, and therefore the treatment of these regions have a small effect on the overall volume evolution.
Obviously, if one considers a more complex geometry where self-interaction dominates the evolving volume, full hydrodynamical simulations are the only reliable tool of investigation.

The Kompaneets approximation itself has some drawbacks, in that it generally predicts too large a velocity, and therefore size, for the shock once it accelerates \citep{Koo90, Matzner99}.
In the context of the present problem, this is more pronounced in the outer density region with a steeper, $R^{-3}$, gradient.
If much of the shock is in the $R^{-3}$ region, the overestimation of velocities and sizes will therefore be greater.

\section{Galactic nuclei model} \label{sec:method_nuclei}
{ In our quiescent SOIs, with no appreciable inflow of gaseous material from further out, the gas density distribution is that of an RIAF.
The distribution of early-type stars (the only population of interest here) is dictated only by local and current conditions, not bearing imprints of the long term history of the assembly of the nucleus.
These facts will allow us to rescale features of our Galactic Centre (observationally constrained because of its proximity) to quiescent nuclei with different SMBH masses.

\subsection{Characteristic radii}
We consider SN explosions within the SOI of an SMBH.
Their fate can be influenced by both the SMBH gravity and its gaseous environment.
Correspondingly, there are characteristic radii in the nucleus associated with these properties.
The first is that of the SOI: the range out to which the gravity of the SMBH dominates over that of the gravitational potential of the bulge.
Following the definition of \cite{Peebles72}, we use
\begin{equation} \label{eq:soi}
R_\mathrm{SOI} \equiv \frac{G\Mbh}{\sigma^2} ,
\end{equation}
for a black hole of mass $\Mbh$, where $\sigma$ is the velocity dispersion of stars about the SMBH.
We use this parameter not only to define the outer edge for the range of explosion distances considered, but also to rescale Milky Way properties to galactic nuclei with different $\Mbh$.
To obtain an expression for the SOI which depends only on $\Mbh$, we use the well-known (`$\Mbh$--$\sigma$') relation between black hole mass and velocity dispersion \citep{Ferrarese00,Gebhardt00}.
Using the observationally determined \cite{Gebhardt00} result,\footnote{Recent studies imply that the $\Mbh$--$\sigma$ relation is steeper than this, and $\sigma$ may have an exponent closer to $5$ \citep[for example,][]{Morabito12}. Although there is still some ambiguity in the value of this exponent, we tested the effect of a very steep relationship $\Mbh = 1.2 \times 10^8 \left( \sigma / \left(200~\mathrm{km \, s}^{-1}\right)\right)^{5.3} \Msun$ motivated by \cite{Morabito12}. Even with this large exponent, we find that our main results, the time-scales in Section~\ref{sec:results_timescales}, are generally only increased by a factor of 2 (while the scaling of radii by the spheres of influence also increases by at most a factor of 2). As the overall consequence is small, we do not present additional results for a steeper $\Mbh$--$\sigma$ relation in this work.} $\Mbh = 1.2 \times 10^8 \left(\sigma / \left(200~\mathrm{km \, s}^{-1}\right)\right)^{15/4} \Msun$, we obtain
\begin{equation} \label{eq:soinosigma}
R_\mathrm{SOI} \approx 3 \left(\frac{\Mbh}{4.3 \times 10^6~\Msun} \right)^{7/15}~\mathrm{pc} ,
\end{equation}
where here, and hereafter, we rescale equations for the Galactic Centre black hole mass.
For what follows, a useful parameter to rescale the Milky Way properties is the ratio $\zeta \propto \Mbh^{7/15}$, between the $R_\mathrm{SOI}$ of a generic $\Mbh$ and that of Sgr A*.

The closer a supernova explodes to the SMBH, the stronger the tidal forces, which may become high enough to disturb and eventually disrupt the remnant in a dynamical time.
This happens when the velocity of the shock front becomes comparable to the Keplerian velocity, $v_\mathrm{K}$, associated to the SMBH gravity field.
We do not model distortions due to tidal effects but we account for the tidal disruption of the remnant when we quantify its `lifetime' (see Section~\ref{sec:results_timescales}).
To this end we test for whether $v_\mathrm{s} < v_\mathrm{K}$ to detect parts of the shock that have decelerated enough to be sheared by the SMBH.
We therefore introduce another characteristic radius---the innermost radius for the existence of SNRs, $R_\mathrm{sh}$, which is limited by SMBH shearing.
This minimal shearing radius is the point at which $v_\mathrm{K}$ is comparable to the initial SNR ejecta velocity, $v_\mathrm{init}$:
\begin{align} \label{eq:shearing_radius}
R_\mathrm{sh} &\equiv \frac{G\Mbh}{v_\mathrm{init}^2} \nonumber\\
              & = 1.9 \times 10^{-4} \left(\frac{\Mbh}{4.3 \times 10^6 \Msun}\right) \left(\frac{v_\mathrm{init}}{10^4~\mathrm{km \, s}^{-1}}\right)^{-2}~\mathrm{pc} \nonumber\\
              & = 900 \left(\frac{v_\mathrm{init}}{10^4~\mathrm{km \, s}^{-1}}\right)^{-2} R_\mathrm{g},
\end{align}
where $R_\mathrm{g}$ is the gravitational radius of the SMBH.

Supernovae that occur {within} $R_\mathrm{sh}$ are completely sheared, as the velocity of the ejecta in all directions is less than the Keplerian velocity around the SMBH.
Note, however, that SNRs can be sheared also at larger radii as the ejecta slows down, and may reach the local Keplerian velocity at a radius larger than $R_\mathrm{sh}$. 
Comparing the shock velocity to the Keplerian velocity is effectively equivalent to comparing the ram pressure with the ambient baryonic pressure, $P_\mathrm{gas}$, since $v_\mathrm{K} \approx c_\mathrm{s} \propto \sqrt{P_\mathrm{gas}/\rho}$ (where $c_\mathrm{s}$ is the sound speed in the external medium).
Additionally, we note that for the same reason we can ignore the shearing of remnants during the initial explosion for $a > R_\mathrm{sh}$, we can also neglect the (Keplerian) orbital motion of the progenitor stars.

Finally, for the gas models, the reference radius for the density, $R_0$, and the location of the break in the gas density power law $R_\mathrm{b}$ (as explained in Fig.~\ref{fig:geometry}) are scaled in our model by the SOI, such that $R_0 \equiv \zeta R_{0,\mathrm{MW}}$ and $R_\mathrm{b} \equiv \zeta R_{\mathrm{b, MW}}$.

\subsection{Gas models}
As mentioned previously, we expect that quiescent SMBHs are surrounded by RIAFs, similar to that which is suggested in the Galactic Centre.
This implies that all flows have similar density gradients, which is set by the physical processes which characterize this accretion regime.
Additionally, their accretion rate must be modest and, in particular, lower than the critical value for advection-dominated accretion of $\dot{M}_\mathrm{crit} = \alpha^2 \dot{M}_\mathrm{Edd}$, or
\begin{equation}
\dot{M}_\mathrm{crit} = 9 \times 10^{-3} \left(\frac{\Mbh}{4.3 \times 10^6~\Msun}\right)~\Msun \, \mathrm{yr}^{-1} ,
\end{equation}
where $\dot{M}_\mathrm{Edd}$ is the Eddington rate, $\alpha = 0.3$ \citep{Narayan95b} and we assume a $10\%$ radiation efficiency for $\dot{M}_\mathrm{Edd}$. 
At a reference distance of $R_{0,\mathrm{MW}} \approx 0.04 ~\mathrm{pc}$, the accretion rate is estimated from observations and simulations to be around $\dot{M} \approx 10^{-5}~\Msun \, \mathrm{yr}^{-1}$ \citep{Cuadra06, Yuan07}.
Therefore, Sgr~A$^*$ is accreting at $\sim 10^{-3}$ of its critical rate.

We extend properties of the Sgr A* accretion flow to other quiescent nuclei as follows.
The primary material for accretion in quiescent nuclei originates from the winds from massive stars in the SOI.
Since the total mass of stars in the SOI scales with $\Mbh$, then so too will the number of massive stars and, therefore, the accreted mass: $\dot{M} \propto \Mbh$.
As above, $\dot{M}_\mathrm{crit} \propto \Mbh$, as well, and therefore the ratio $\dot{M}/\dot{M}_\mathrm{crit}$ is constant over $\Mbh$.
In other words, for our physically motivated picture of ‘quiescent’ nuclei, the SMBH is accreting at the same fraction of Eddington as Sgr A* ($\dot{M} \approx 10^{-5} \dot{M}_\mathrm{Edd}$).\footnote{It is possible for SMBHs to be accreting at different fractions $\dot{M}/\dot{M}_\mathrm{crit} < 1$ and still be termed `quiescent' in the conventional sense. However, the accretion rate given here is the most physically motivated value based on scaling of quantities by $\Mbh$, and deviations from this value are beyond the scope of this work.}

Given this accretion rate of $\dot{M}/\dot{M}_\mathrm{crit} \approx 10^{-3}$, we can estimate the density at $R_0$ for nuclei with a different $\Mbh$.
To do so we use the continuity equation for the flow,
\begin{equation}
\dot{M} \approx 4 \uppi R_0^2 \, \rho(R_0) \, v_\mathrm{K}(R_0) ,
\end{equation}
where the scale height $H \approx R$ and the radial velocity $v_\mathrm{R} \approx v_\mathrm{K}$.

The reference density for our general galactic nuclei models is therefore
\begin{equation} \label{eq:extragal-numdensity}
n_0 = n(R_0) \approx 130 \left( \frac{\Mbh}{4.3 \times 10^6 \Msun} \right)^{1/2} \zeta^{-3/2}~\mathrm{cm^{-3}} .
\end{equation}
An example of our density model for $\Mbh = 10^7 \Msun$ and three different inner gradients is given in Fig.~\ref{fig:density_example}, where the effect on the density profiles of fixing the scaling reference point at $R_0$ is evident.
\begin{figure}
\begin{center}
\includegraphics[width=\columnwidth]{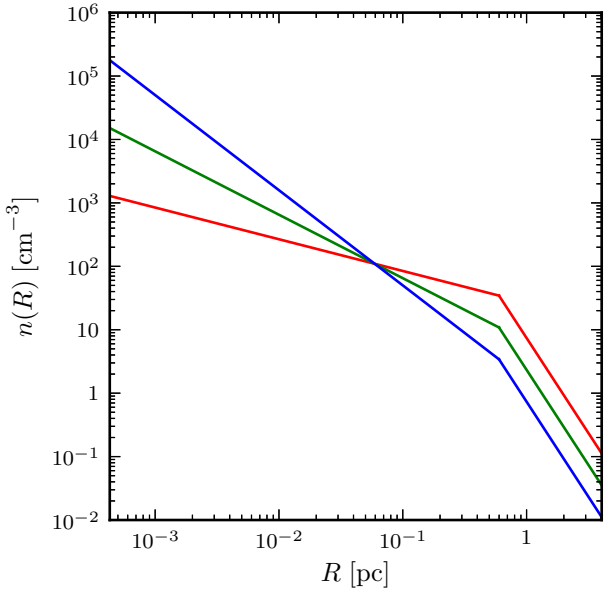}
\end{center}
\caption{Example of a gas density model used for a galaxy with a $10^7 \Msun$ SMBH, showing the number density, $n$, as a function of radius $R$ from the SMBH.
The red (shallowest) line corresponds to an inner gradient with $\omega_\mathrm{in} = 1/2$, the green to $\omega_\mathrm{in} = 1$ and the blue (steepest) to $\omega_\mathrm{in} = 3/2$.
All models have a gradient outside the break of $\omega_\mathrm{out} = 3$.
The density is scaled using a reference point $R_0$, seen as the point of convergence of all the inner density gradients (in this case, $R_0 = 0.06~\mathrm{pc}$).
A break in the density distribution is located at a constant $R = R_\mathrm{b}$ for all choices of the density gradient (in this case, $R_\mathrm{b} = 0.6~\mathrm{pc}$).
The left-hand and right-hand limits of the horizontal axis are determined by the shearing radius (equation \ref{eq:shearing_radius}) and SOI (equation \ref{eq:soi}), respectively.}
\label{fig:density_example}
\end{figure}

\subsection{Massive star distributions} \label{sec:method_massive_star_distributions}
Given a physical number density $n_*(R)$ of stars at a distance $R$ from the centre of mass, the projection on to the celestial sphere gives, as a function of the projected radius $R_\mathrm{pr}$, a surface density of stars $\Sigma_*(R_\mathrm{pr})$.
Observationally, the latter quantity is typically given.
Assuming a spherically symmetric spatial distribution, it is possible to reverse the projection to infer the spherical number density,
\begin{equation}
n_*(R) = \frac{-1}{\uppi} \int_{R}^{\infty}{\frac{\ud \Sigma_* \left(R_\mathrm{pr}\right)}{\ud R_\mathrm{pr}} \frac{\ud R_\mathrm{pr}}{\sqrt{R_\mathrm{pr}^2 - R^2}}} ,
\end{equation}
provided that the physical number density $n_*(R)$ falls off at large $R$ at a rate greater than $R^{-1}$.
For power-law distributions, this gives a correspondence of the observed radial dependence, $R_\mathrm{pr}^{-\Gamma}$, to the physical dependence, $R^{-\gamma}$, via the relationship $\gamma \sim \Gamma +1$.

For the Milky Way, there is evidence for two different power-law distributions in the old and young stellar populations of the Galactic Centre.
A particular curiosity is an apparent depletion of late-type (K, M) giants in the inner $0.5~\mathrm{pc}$ \citep{Do09}.
This leads to a much shallower (possibly inverted) inner power-law for the late-type distribution compared to that of the early-type (O, B) stars.
Recent analyses estimate the radial dependence of the early-type stars in the Milky Way nuclear star cluster to be approximately $R_\mathrm{pr}^{-1}$ inside the power-law  break and $R_\mathrm{pr}^{-3.5}$ outside \citep{Buchholz09, Do13a}, corresponding to values of $\gamma$ of $2$ and $4.5$, respectively.

In the nuclei of different galaxy types, variations in the stellar distributions for longer-lived stars are possible due to differing nuclear assembly histories.
However, in our picture of a self-regulating SOI, the young star distributions are taken to be the same across the range of $\Mbh$, where the most recent star formation in this region is indifferent to the history of the nucleus.
Therefore, for our galactic nuclei model, we use the same values for $\gamma$ as those given above for the early-type stars around Sgr A*.
As before, we scale the break in the stellar number density by the SOI of the SMBH, which defines the transition radius between the two values of $\gamma$.

\section{Results} \label{sec:results}
We proceed to describe our main results, based on the method outlined in the previous sections.
In Section~\ref{sec:results_deceleration_lengths}, we examine the effect of black hole mass and gas density profile on the deceleration length using the prescriptions of Section~\ref{sec:method_deceleration_lengths}.
Then, using the numerical method of Section~\ref{sec:method_numerical}, the overall SNR morphology is presented in Section~\ref{sec:results_morphologies}.
Finally, in Section~\ref{sec:results_timescales}, we investigate the X-ray emitting lifetimes based on shearing of the SNR ejecta by the SMBH.
We then use this last result to predict the mean SNR lifetimes expected within the SMBH spheres of influence for different $\Mbh$.

\subsection{Deceleration lengths} \label{sec:results_deceleration_lengths}
\begin{figure*}
\centering
\includegraphics[width=2.1\columnwidth]{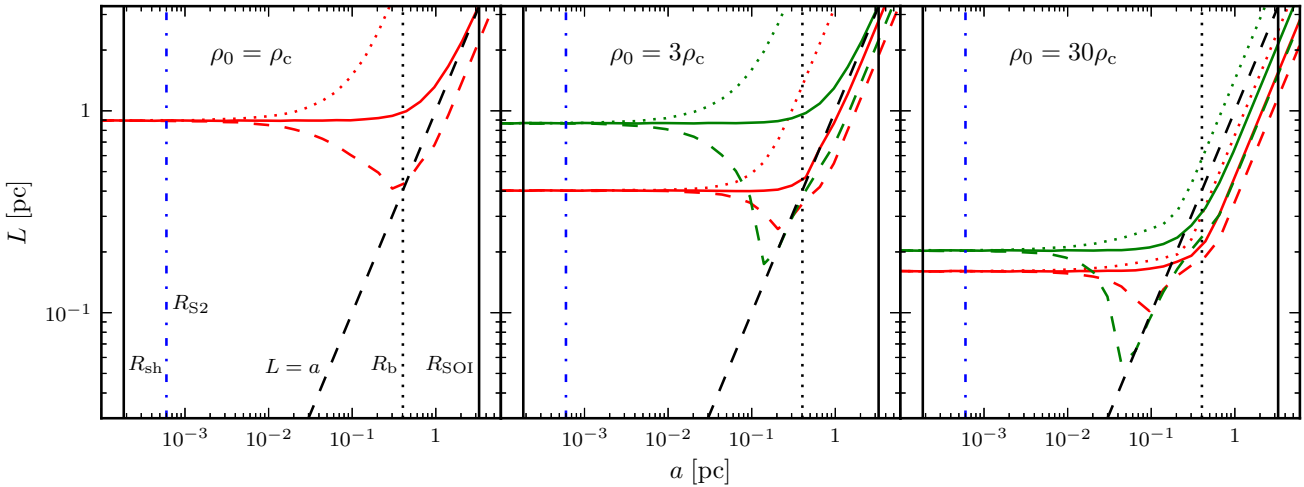}
\caption{Deceleration lengths, $L$, as a function of explosion distance from the SMBH, $a$, for a Milky Way model ($\Mbh = 4.3 \times 10^6 \Msun$) with density scalings of the canonical density $130~\mathrm{cm}^{-3}$ of $1$, $3$ and $30$ times (left to right).
Red lines show the $\omega_\mathrm{in} = 1/2$ model and green lines show the $\omega_\mathrm{in} = 1$ model.
The solid curves are the average deceleration lengths derived by integrating the mass over a sphere, as discussed in Section~\ref{sec:method_deceleration_lengths}.
The dotted and dashed curves are integrals away from and towards the origin, respectively.
The shearing radius ($2 \times 10^{-4}~\mathrm{pc}$) and the SOI for Sgr A* are marked by $R_\mathrm{sh}$ and $R_\mathrm{SOI}$, respectively.
The vertical dotted black line is the location of the break in density, outside of which $\omega_\mathrm{out} = 3$.
The diagonal dashed black line shows $L = a$, which represents the distance to the SMBH.
The blue dot--dashed line shows the pericentre distance of the star S2.
The curves for $\omega_\mathrm{in} = 1$ are not plotted in the left-hand panel since most of the points lie outside of the SOI axis bounds.}
\label{fig:deceleration_lengths_MW}
\end{figure*}

The SNR begins to appreciably decelerate once the swept up mass becomes comparable to the ejecta mass.
This end of the ejecta-dominated stage can be characterized by a deceleration length from the explosion point, $L$, determined by the density integrals of Section~\ref{sec:method_deceleration_lengths}, which varies with direction.
This deceleration length depends on the gas density and on the radial density profile.
These depend respectively on the SMBH accretion rate and accretion mode (for example, CDAF versus ADAF).

Considering the values of $L$ in various environments, we obtain an indication of the length- and time-scales over which SNRs will end their ejecta-dominated stage and start  decelerating.
As a reference for the crossing time-scale of a nucleus, recall that an SNR that does not appreciably decelerate from its initial $\sim 10^4~\mathrm{km \, s}^{-1}$ would reach a radius of $1~\mathrm{pc}$ in approximately $100~\mathrm{yr}$.
An investigation of $L$ in different directions indicates which SNRs will decelerate within this time-scale. 
It also provides a test for the level of asymmetry of the SNR during this stage of evolution.
We will later proceed to model SNRs through the decelerating stage.

Figs~\ref{fig:deceleration_lengths_MW} and \ref{fig:deceleration_lengths} depict two curves describing the deceleration length, $L$, approximately towards and away from the SMBH.
The angle $\psi_0$ approximately towards the SMBH is taken to be $10^{-3}~\uppi$, such that the integrated path through the density runs very close to the SMBH, but does not pass though the singularity at the origin.
The difference between these two curves provides a measure of the asymmetry of the remnant at the end of the ejecta-dominated stage.
For comparison, in Fig.~\ref{fig:deceleration_lengths_MW}, a third curve (solid line) is shown that describes an average deceleration length derived by an integral over a sphere.

We first consider a model of the background gas in the Milky Way.
Fig.~\ref{fig:deceleration_lengths_MW} shows $L$ for three different density values: the observationally motivated `canonical' density $\rho_0 = \rho_\mathrm{c}$ (in number density, $n_\mathrm{c} \approx 130~\mathrm{cm}^{-3}$, left-hand panel), and  $3$ (central panel) and $30$ (right panel) times that value.
We consider values other than the canonical density, as, even in quiescent nuclei such as the Galactic Centre, there is the possibility for variation in the overall density of the accretion flow.
For example, denser accretion flows can result from sudden accretion episodes from tidally disrupted stars or clouds,\footnote{A recent example around Sgr A* is the object G2 \citep[for example,][]{Burkert12}; though, if a cloud, its mass is too small to have a significant impact on the overall density.}
or they can be associated with more intense star formation activity in the nucleus.
Scaling the density also shows the effect of under- or misestimating the gas density from the X-ray emission.

As expected with increasing density, there is an overall trend towards lower values of $L/a$. 
There is also a trend towards more symmetric remnants with increasing density, since in general the ratio $L/a$ is reduced for higher densities.

The investigation of different density profiles (see Fig.~\ref{fig:density_example}) leads us to conclude that CDAF/ADIOS model, preferred by Galactic Centre observations \citep{Wang13}, gives, quite generally, shorter deceleration lengths (red lines in Figures \ref{fig:density_example} and \ref{fig:deceleration_lengths_MW}).
The flatter CDAF/ADIOS profile (smaller $\omega_\mathrm{in}$) is denser in most of the SOI of the black hole, therefore reducing $L/a$.

For the canonical value of density in the Milky Way (left panel of Fig.~\ref{fig:deceleration_lengths_MW}), the deceleration lengths are $L \gtrsim 1~\mathrm{pc}$.
Considering the CDAF model, we remark that, for the canonical density, the majority of the SNRs would decelerate beyond the SMBH location.
Ejecta from a star such as S2 (marked with a blue dot--dashed line in Fig.~\ref{fig:deceleration_lengths_MW}) is expected to evolve more symmetrically than that from a star further out, in the stellar disc(s) ($\sim R_\mathrm{b}$).
Already with a factor of few enhancement in density, SNRs in and beyond the stellar disc would decelerate appreciably before they reach Sgr A* (see central panel). 
\begin{figure*}
\captionsetup[subfigure]{labelformat=empty}
\centering
\subfloat[]{
\includegraphics[width=\columnwidth]{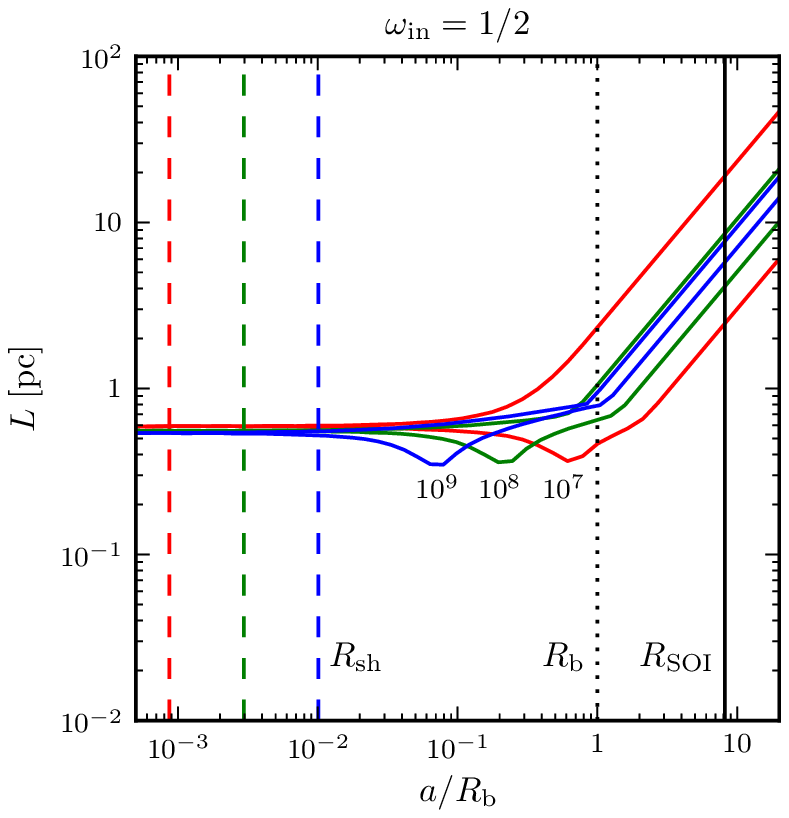}
}
\subfloat[]{
\includegraphics[width=\columnwidth]{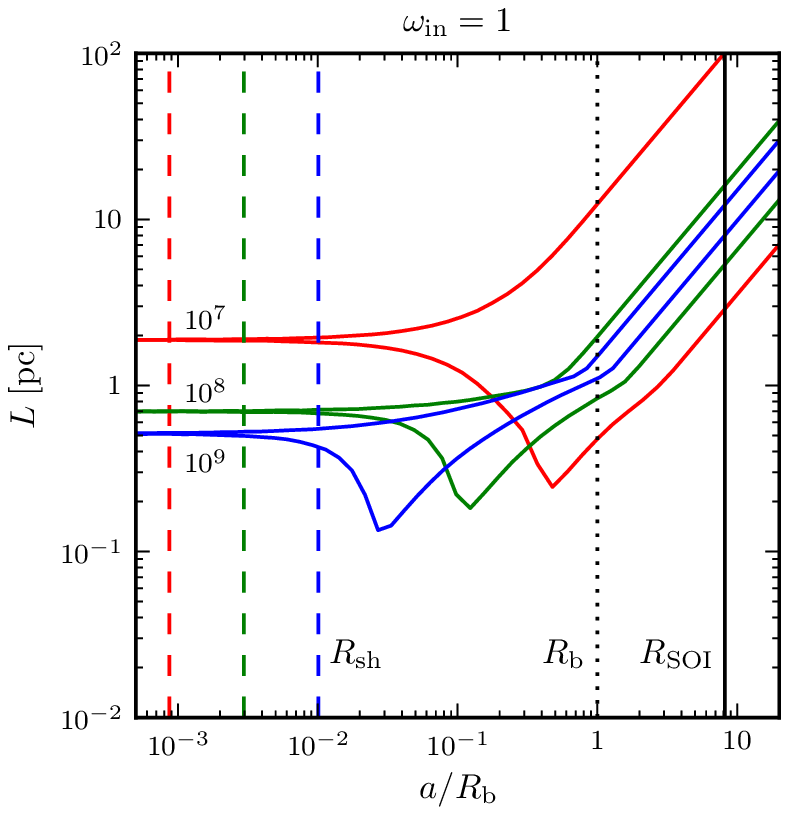}
}
\vfill
\subfloat[]{
\includegraphics[width=\columnwidth]{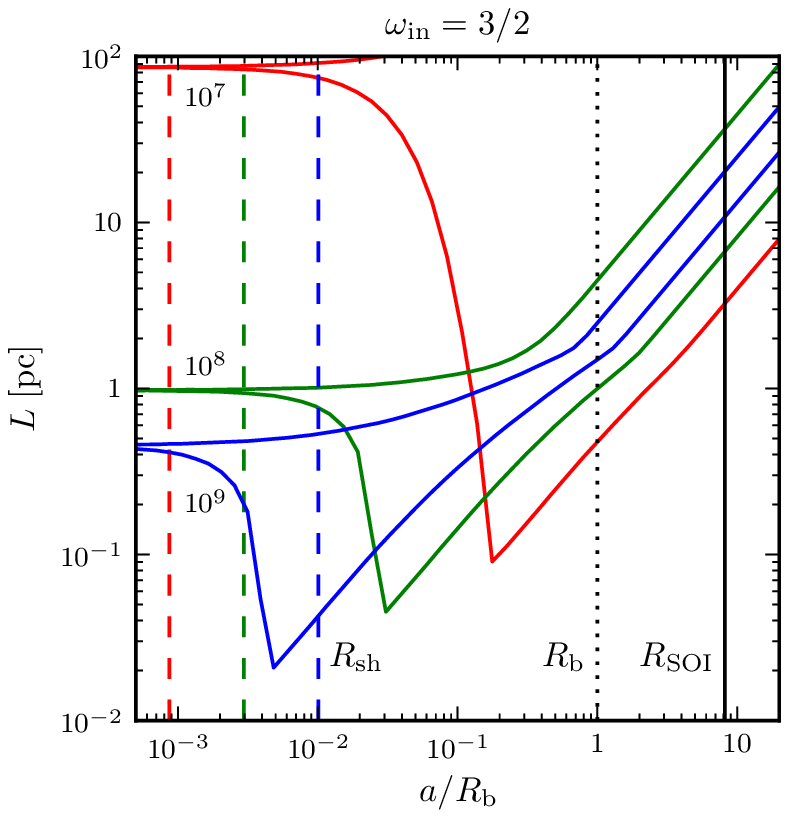}
}
\caption{Deceleration lengths, $L$, as a function of the ratio of the explosion distance, $a$, to the break in gas density, $R_\mathrm{b}$.
Each set of axes is to the same scale, and corresponds to a different gradient near the SMBH: $\omega_\mathrm{in} = 1/2$ (top left), $\omega_\mathrm{in} = 1$ (top right) and $\omega_\mathrm{in} = 3/2$ (bottom).
As indicated in the top-left panel, red corresponds to the gaseous environment of an SMBH of $10^{7}~\Msun$, green to $10^{8}~\Msun$ and blue to $10^{9}~\Msun$.
Two curves are shown for each $\Mbh$ (each colour); the higher curve shows $L$ for the direction away from the SMBH, and the lower curve shows $L$ towards the SMBH.
The dashed coloured lines correspond to $R_\mathrm{sh}$ for each value of $\Mbh$.
The dotted black line shows the break in gas density at $R_\mathrm{b}$, while the solid black line shows the extent of the SOI (which is the same multiple of $R_\mathrm{b}$ for all \Mbh).
Unlike in Fig.~\ref{fig:deceleration_lengths_MW}, we also show results for a standard ADAF model ($\omega_\mathrm{in} = 3/2$; bottom panel), since these density profiles can produce $L$ values that fall within $R_\mathrm{SOI}$.\label{fig:deceleration_lengths}}
\end{figure*}

Fig.~\ref{fig:deceleration_lengths} shows deceleration lengths for each of $\omega_\mathrm{in} \in \left\{1/2, \, 1, \, 3/2 \right\}$ for galactic nuclei with SMBH masses of $10^{7}~\Msun$ (red), $10^{8}~\Msun$ (green) and $10^{9}~\Msun$ (blue).
Hereafter, we scale the explosion distance by the location of the break in gas density, $R_\mathrm{b}$, as it is the value of $a$ at which a change in behaviour of SNRs is expected.
Accretion rates (and thus gas density normalization) and characteristic radii are rescaled as explained in Section~\ref{sec:method_nuclei}, and all increase with black hole mass.

For $\Mbh \lesssim 10^8~\Msun$, the density is low enough and $R_\mathrm{sh}$ is small enough that for small values of $a$ the SNR can expand over the SMBH before appreciably decelerating.
The centre of such a shock front is close to being aligned with the centre of symmetry of the gas distribution, which leads to a more symmetric evolution.
This is seen in the fact that all $L$ values converge at small $a$.
However, by $\Mbh \gtrsim 10^9~\Msun$, the gas density is high enough and $R_\mathrm{sh}$ extends far enough from the SMBH that explosions near the SMBH cannot pass over the singularity before being sheared, and this more symmetric expansion regime beyond the SMBH at small $a$ is no longer present.

In general, with increasing overall density (around more massive black holes), we find the same trend observed for the Milky Way with denser gaseous environments: the ratio of $L/a$ decreases, as does the maximum possible asymmetry (differences between upper and lower curves).
On the other hand, increasing $\omega_\mathrm{in}$ also tends to create greater asymmetries in the SNRs, as can be appreciated by comparing the three panels of Fig.~\ref{fig:deceleration_lengths}.
For higher $\omega_\mathrm{in}$, the ratio $L/a$ is higher in the direction away from the SMBH as a result of the density being lower at any point further than $R_0$, but it is lower towards the SMBH as the SNR sweeps through a steeper density gradient near the origin.

\subsection{Morphological evolution} \label{sec:results_morphologies}
We turn now, using the numerical treatment of Section~\ref{sec:method_numerical}, to the subsequent adiabatically decelerating evolution of the remnant.
We consider explosions both inside and outside the density break, $R_\mathrm{b}$.
The explosion distances are chosen such that, across all $\Mbh$, the same $a/R_\mathrm{b}$ ratio is maintained for the examples inside $R_\mathrm{b}$ ($a/R_\mathrm{b} = 0.6$) and outside $R_\mathrm{b}$ ($a/R_\mathrm{b} = 2.5$); this will also be useful for comparison with the time-scale plots of Fig.~\ref{fig:lifetimes}.

Fig.~\ref{fig:morphologies_MW} depicts the morphology for two explosion distances in the Milky Way environment.
The explosion inside the density break does not decelerate before reaching the SMBH (as expected from Fig.~\ref{fig:deceleration_lengths_MW}, where $L > a$).
Therefore it passes over the SMBH without any significant distortion. 
The SNR subsequently expands into the lower density region almost spherically, with the centre of the SNR being very near the SMBH.
For the explosion outside the break in density, the trailing part of the shock decelerates before reaching the SMBH, while at the same time the parts expanding through the $\omega = 3$ region wrap around the SMBH and eventually self-interact.
SNRs such as this, which explode far enough from the SMBH that $L < a$, show significant asymmetries during their evolution.
\begin{figure*}
\begin{center}
\includegraphics[width=1.8\columnwidth]{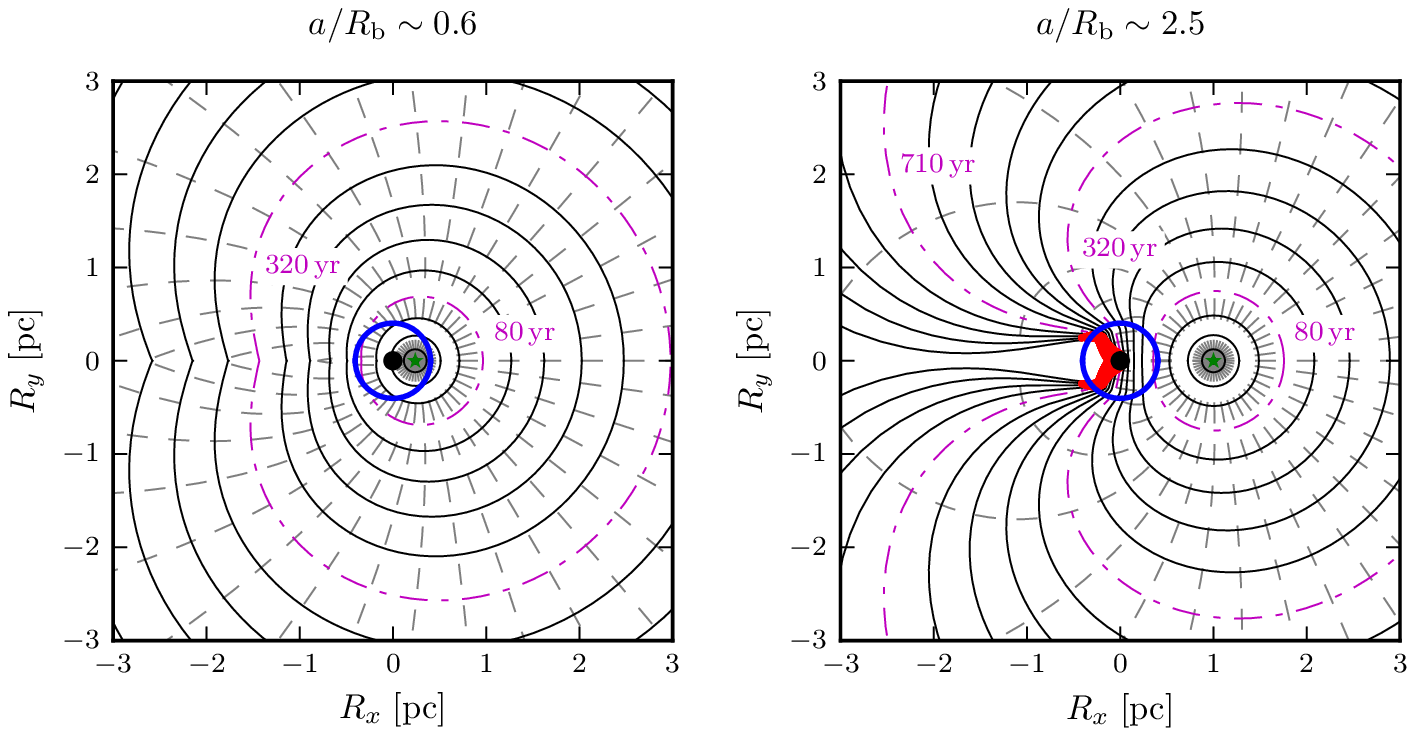}
\caption{Morphological evolution for two explosion distances for the Milky Way, inside (left) and outside (right) $R_\mathrm{b}$.
The explosion distance in the left-hand panel is $a = 0.24~\mathrm{pc}$, and in the right-hand panel, $a = 1.0~\mathrm{pc}$.
The black hole is marked with a black point at the origin and the supernova occurs at the green star.
The break in density at $R_\mathrm{b}$ is shown as a thick blue circle.
Snapshots of the shock front at different times are in solid black, and dashed lines show sample flowline paths.
Flowlines flagged as having reached the shearing condition discussed in Section~\ref{sec:results_timescales} ($v_\mathrm{SNR} < v_\mathrm{K}$) are shown as red points.
Some example snapshot times are indicated next to corresponding dot--dashed magenta lines.
In the left-hand panel, the initial snapshot shown is at $13~\mathrm{yr}$ and the final at $620~\mathrm{yr}$; in the right panel, the initial snapshot is at $13~\mathrm{yr}$ and the final at $1100~\mathrm{yr}$.
The spacing between snapshots is an equal multiple of the (variable) time step used.
In the left-hand panel, the initial snapshot spacing is $16~\mathrm{yr}$, and the final spacing is $84~\mathrm{yr}$; in the right-hand panel, the initial snapshot spacing is $16~\mathrm{yr}$, and the final spacing is $120~\mathrm{yr}$.
In both cases, the remnants expand well beyond the window shown, until they reach the radiative stage as discussed in Section~\ref{sec:results_radiative}.}
\label{fig:morphologies_MW}
\end{center}
\end{figure*}

\begin{figure*}
\begin{center}
\includegraphics[width=2\columnwidth]{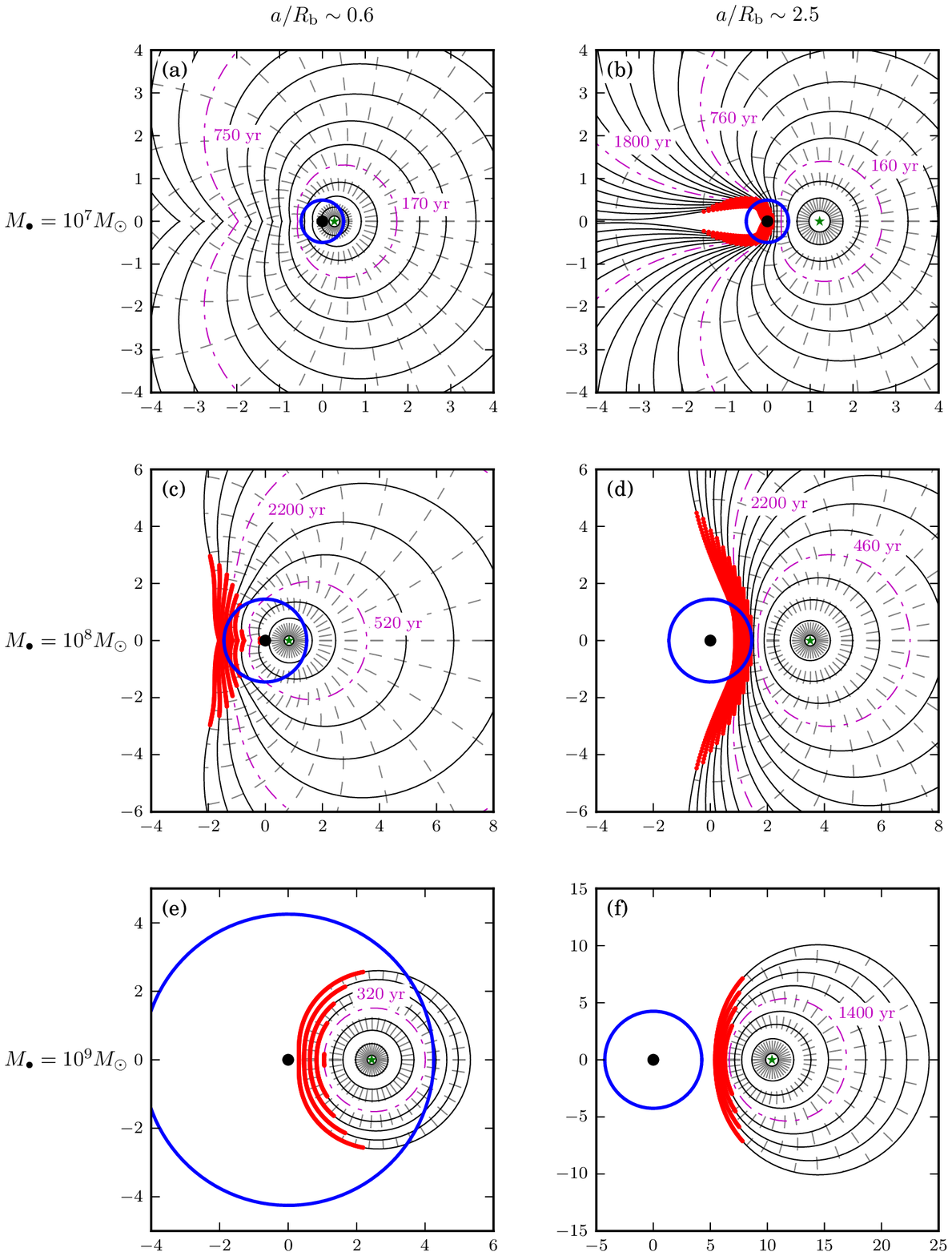}
\caption{Example remnant morphologies for three black hole masses in an $\omega_\mathrm{in} = 1/2$ density for explosions at two $a/R_\mathrm{b}$ ratios (also indicated as dotted lines on Fig.~\ref{fig:lifetimes}).
All axes are in units of parsecs.
The top row (panels `a' and `b') shows $\Mbh$ of $10^7~\Msun$, the middle (`c' and `d') $10^8~\Msun$ and the bottom (`e' and `f') $10^9~\Msun$.
Specific details of the parameters in each panel are given in Table \ref{tab:morphologies}.
All markers, line styles and colours are as for Fig.~\ref{fig:morphologies_MW}.
Snapshots are plotted up until the end of the life of the remnant discussed in Section~\ref{sec:results_timescales} (if found).}
\label{fig:morphologies}
\end{center}
\end{figure*}
Fig.~\ref{fig:morphologies} shows examples of the variation in morphology of the remnant arising from differences in the black hole mass (and therefore the gas density).
The values of $\Mbh$ are the same as those used in Fig.~\ref{fig:deceleration_lengths} for the deceleration lengths; the black hole mass increases from top to bottom.
The data shown in Fig.~\ref{fig:morphologies} are summarized in Table \ref{tab:morphologies}.

\begin{table*}
\centering
\begin{tabular}{ccccccc}
 panel & $\Mbh~(\Msun)$ & $a~(\mathrm{pc})$ & $t_\mathrm{init}~(\mathrm{yr})$ & $\Delta t_\mathrm{init}~(\mathrm{yr})$ & $\Delta t_\mathrm{final}~(\mathrm{yr})$ & $t_\mathrm{final}~(\mathrm{yr})$ \\
\hline
a & $1 \times 10^7$ & 0.28 & 14 & 23  & 160  & $1200 \, \dagger$\\
b & $1 \times 10^7$ & 1.2  & 28 & 32  & 1000 & $3300 \, \dagger$\\
c & $1 \times 10^8$ & 0.83 & 19 & 91  & 760  & $4300$\\
d & $1 \times 10^8$ & 3.5  & 24 & 58  & 580  & $5100$\\
e & $1 \times 10^9$ & 2.4  & 16 & 40  & 210  & $1000$\\
f & $1 \times 10^9$ & 10   & 70 & 170 & 900  & $4300$
\end{tabular}
\caption{Parameters used for the plots of Fig.~\ref{fig:morphologies}, where $\Mbh$ is the mass of the SMBH and $a$ is the distance of the explosion point from the SMBH.
The snapshots that are plotted are at fixed multiples of the time step, although the time step is dynamic and increases with the size of the remnant.
The first curve that is plotted is at a time $t_\mathrm{init}$, and the difference between the first two snapshots is given in $\Delta t_\mathrm{init}$.
The difference between the final two snapshots is given in $\Delta t_\mathrm{final}$, and the last curve that is plotted is at a time $t_\mathrm{final}$.
Additional examples of snapshot times are shown on Fig.~\ref{fig:morphologies} with corresponding magenta dot--dashed curves.
Note that the values of $\Delta t$ are {not} the time step used by the numerical treatment, which is much smaller.
$\dagger$: simulation ran until $10^4~\mathrm{yr}$, stopping due to the radiative onset threshold discussed in Section~\ref{sec:results_timescales}, but the last curve shown within these axis limits is at the stated time.}
\label{tab:morphologies}
\end{table*}

It is evident that, in many cases, much of the mass of the remnant remains near the SMBH due to the focusing effect of the density gradient on the flowlines.
Therefore, unlike expansion away from the SMBH, where the mass behind the shock is more tenuous due to the rapid shock expansion, the ejecta material near the SMBH is expected to be more concentrated.

In addition to the symmetries in $L$ found in Section~\ref{sec:results_deceleration_lengths}, explosions closer to a lower-mass SMBH ($\Mbh \lesssim 10^8~\Msun$) are also found to be more symmetric during their adiabatic evolution compared to those further from the SMBH (compare panel `a' of Fig.~\ref{fig:morphologies} to panel `b').
SNRs near lower $\Mbh$ expand over the SMBH largely unimpeded, and their centres are closely aligned with the centre of symmetry of the gas distribution when they enter the adiabatic phase.

Remnants around higher-mass SMBHs ($\Mbh \gtrsim 10^9~\Msun$) are more symmetric for the whole duration of their adiabatic lifetimes.
The overall increase in density causes the scale of the remnant to be small (relative to the scale of the background gas distribution), and so, within the lifetime of the remnant, significant asymmetries have not yet developed.
Indeed, by looking at the overall remnant size with respect to $\Mbh$, there is a clear trend towards decreasing remnant size with increasing $\Mbh$ during this phase of evolution.

\subsection{Adiabatic SNR lifetimes} \label{sec:results_timescales}
During the adiabatic phase, the SNR can be observed as a hot, X-ray emitting object.
The hard X-rays can penetrate the obscuring matter in galactic nuclei and allow us to detect the SNR.
In particular, in quiescent nuclei, the SMBH light may not prevent the SNR detection in X-rays.
We will refer to this temporal window during which the SNR can be observed in X-rays as the ejecta `adiabatic lifetime' or simply its `lifetime', because once the ejecta become radiative it cools rapidly and its X-ray emission ceases. 

As previously mentioned, in a general environment, the adiabatic phase ends when the expansion has caused the internal temperature to decrease enough for radiative losses to become dynamically important.
In the SOI, however, the tidal field can tear apart the ejecta well before the end of its radiative phase.
In this case, the ejecta is dispersed and may not be immediately identified as such, at any wavelengths.

So far, we have been evolving the shock expansion without directly considering the gravitational 
field of the SMBH, but only considering its indirect influence in shaping the gas density profiles.
This is a very good approximation as long as the internal pressure forces are significantly larger than the SMBH tidal forces.
From momentum conservation, this means that we can ignore the SMBH gravitational field whenever the shock front is faster than the local Keplerian velocity, $v_\mathrm{K}$.
When, however, $v_\mathrm{s} < v_\mathrm{K}$, the dynamics of the shock front is dictated by gravity.
As different adjacent parts move at different speeds $\sim v_\mathrm{K} \propto R^{-1/2}$, the SNR is sheared in a dynamical time.

If the overall density is large enough, the swept up mass causes the shock front to decelerate before being sheared.
As shown in Sections \ref{sec:results_deceleration_lengths} and \ref{sec:results_morphologies}, most of the ejected mass is focused close to the SMBH (see Fig.~\ref{fig:morphologies}), in regions of high Keplerian velocities, where this is more likely to occur.
This effect is more prominent in the surrounding of more massive black holes.

Practically, we consider that the SNR has ended its life when half of the original ejecta mass satisfies the condition $v_\mathrm{s} < v_\mathrm{K}$.
A fraction of the ejecta mass is assigned to each flowline (as per Section~\ref{sec:method_numerical_scheme}), and we monitor the total proportion of sheared mass $M_\mathrm{sh}$ to determine this time.
The sheared portions of the remnants are marked with red points over the flowlines in Fig.~\ref{fig:morphologies}.
Fig.~\ref{fig:lifetimes} describes the variation in lifetime for different explosion distances and $\Mbh$.
For models with $M_\bullet \lesssim 10^7 \Msun$, no SNR within the SOI ends its lifetime by shearing within the $10^4~\mathrm{yr}$ shown on the plot.
\begin{figure*}
\begin{center}
\includegraphics[width=2.1\columnwidth]{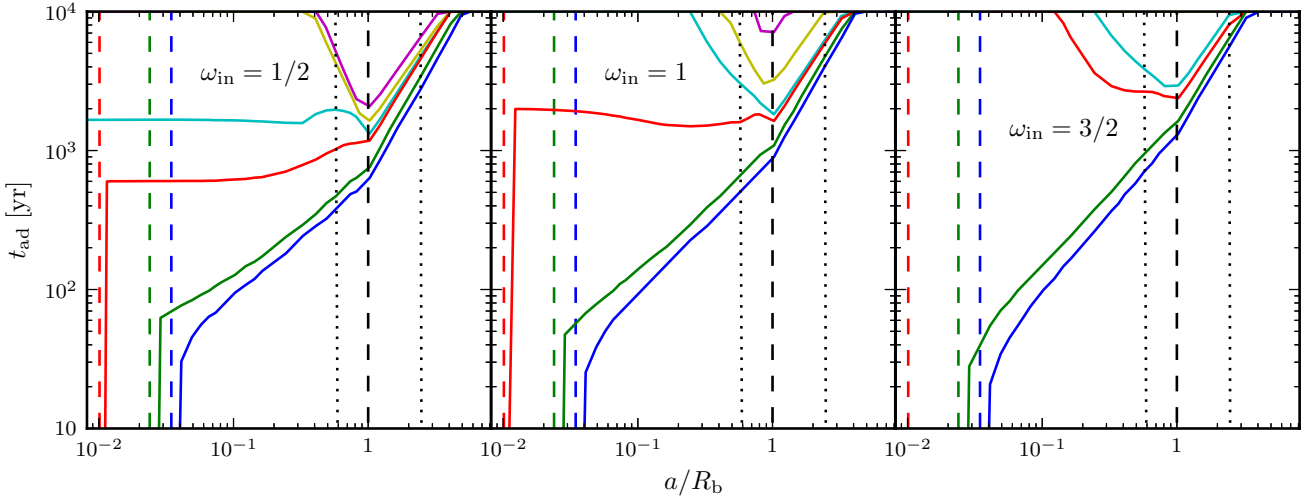}
\caption{Time taken to shear a total ejecta mass $M_\mathrm{sh} \geq M_\mathrm{ej}/2$ for a range of $\Mbh$ and $\omega_\mathrm{in}$, for explosion distances $a$ scaled to the break in gas density, $R_\mathrm{b}$.
{Decreasing $t_\mathrm{ad}$ curves correspond to increasing $\Mbh$}, and the black hole masses shown are as follows; magenta: $5 \times 10^7~\Msun$, yellow: $1 \times 10^8~\Msun$, cyan: $5 \times 10^8~\Msun$, red: $1 \times 10^9~\Msun$, green: $5 \times 10^9~\Msun$, blue: $1 \times 10^{10}~\Msun$.
The curve for each $\Mbh$ represents data from simulations covering between 30 and 40 values of $a$.
Dashed, vertical coloured lines show the corresponding $R_\mathrm{sh}$.
The dashed black line shows the location of $R_\mathrm{b}$.
The dotted black lines show the two regions of explosion distances investigated in Fig.~\ref{fig:morphologies}.
The right-hand limit of the horizontal axis is set at the SOI.
By $10^4~\mathrm{yr}$, radiative losses become significant in all cases; if the shearing condition $v_\mathrm{sh} < v_\mathrm{K}$ has not yet been met by this time, the adiabatic stage ends due to the radiative transition discussed in Section~\ref{sec:results_radiative}.
Note that the Galactic Centre SMBH is not shown on this plot, as we do not find $M_\mathrm{sh} \geq M_\mathrm{ej}/2$ at any value of $a$ before the radiative stage sets in.}
\label{fig:lifetimes}
\end{center}
\end{figure*}

There is a clear trend in the behaviour of the remnant lifetimes as $a$ increases.
The lifetimes for SNRs exploding within $R_\mathrm{sh}$ vanish, by the definition of the shearing radius.
Just outside $R_\mathrm{sh}$ the lifetimes are short but they  increase rapidly with $a$.
This is because, at small $a$, the SNR expands at $v_\mathrm{init} \approx 10^4~\mathrm{km \, s}^{-1}$ for the time that it is near the SMBH.
Therefore, the amount of sheared ejecta is directly proportional to the fraction of the surface area of the SNR that enters the sphere of radius $R_\mathrm{sh}$ around the SMBH.
This quickly decreases as $a$ increases, causing the rapid increase in lifetime.

For $\Mbh \lesssim 10^8~\Msun$, there is a sudden jump to very long lifetimes at distances slightly larger than $R_\mathrm{sh}$ (in fact, within a range of small $a$, the $M_\mathrm{sh} \geq M_\mathrm{ej}/2$ condition is never met before the radiative stage sets in). 
Explosions at small $a$ expand over the SMBH without any significant disruption due to their high initial velocity.
They then almost entirely travel `downhill' in density, and so $\geq 50\%$ of the remnant is never sufficiently decelerated by travelling into a sufficiently high density.\footnote{The amount of deceleration in such cases may be underpredicted by the Kompaneets approximation, as discussed in Section~\ref{sec:method_numerical}.}

Further increasing $a$, the lifetime then drops significantly. This can be attributed to the aforementioned focusing of ejecta towards the SMBH (as seen in Fig.~\ref{fig:morphologies}).
Since the shock decelerates significantly in these same regions, a drop in the lifetime is seen, particularly near the break in gas density, $R_\mathrm{b}$, in Fig.~\ref{fig:lifetimes}.
For lower-mass SMBHs like Sgr A* ($\Mbh \lesssim 10^7~\Msun$), although $50\%$ of the SNR is not found to be sheared, parts of SNRs at this region of $a$ may be decelerated enough before reaching the SMBH that at least a small fraction ($\lesssim 20\%$) of the remnant is sheared.
At large $a$ values, much of the SNR spreads out to distances further from the SMBH before appreciably decelerating.
Here, a combination of lower $v_\mathrm{K}$ and less deceleration of the shock conspire to lengthen the time taken to reach the shearing condition.

Inspection of the left-middle panel of Fig.~\ref{fig:morphologies} shows an intermediate behaviour around $\Mbh \approx10^8~\Msun$.
In this case, for explosions near the SMBH, the SNR can first pass beyond the SMBH and then become significantly sheared.
However, once the SMBH mass is high enough ($\Mbh \gtrsim 5 \times 10^8~\Msun$), the higher densities and larger $v_\mathrm{K}$ mean that it is impossible to have large lifetimes for small $a$, as remnants are always significantly sheared before they expand far from the SMBH.
As seen in Sections \ref{sec:results_deceleration_lengths} and \ref{sec:results_morphologies}, they therefore never enter the more symmetric, `downhill' expansion regime seen for lower values of $\Mbh$.
Therefore, around these higher mass SMBHs, the lifetime is generally $\sim 10^3~\mathrm{yr}$ or less in the entire inner density region.

We note that the most likely distances for core-collapse supernovae to occur are at or within $R_\mathrm{b}$, as the density of massive stars falls much more steeply beyond this radius (see Section~\ref{sec:results_mean_lifetimes}). 
The SNR lifetimes are shortest in this region (see Fig.~\ref{fig:lifetimes}).
For more massive SMBHs the lifetime is also reduced significantly over the whole inner region.

Before concluding this section we note that the tidal shearing of an SNR might have a significant observational signature.
One can expect that  the sheared ejecta will be largely accreted by the SMBH. 
This will temporarily enhance the accretion rate, leading to a period of about 100 years in which a few solar masses are accreted by the black hole. 
This is a very large accretion rate that may lead to a flaring of the SMBH, reaching luminosities comparable to the Eddington luminosity over this period.

\subsubsection{The radiative transition} \label{sec:results_radiative}
As we are interested in the hot, X-ray stage of evolution, we test for the onset of increased radiative losses in the remnant.
This was estimated by comparing the shock velocity to the threshold of $300~\mathrm{km \, s}^{-1}$ outlined in Section~\ref{sec:method_radiative}.
As with the shearing condition, we consider the adiabatic lifetime to have ended when more than half of the remnant is below this velocity.

For SNRs that are not destroyed by shearing, we find that this criterion is not satisfied before the SNR grows larger than the SOI.
However, the density gradient will not continue to have a $R^{-3}$ form indefinitely.
Beyond the SOI, the SMBH no longer has any substantial influence on the environment and the density is expected to level off.

Therefore, to determine the onset of the radiative stage, we find the time when more than half of the shock reaches $300~\mathrm{km \, s}^{-1}$ by extending the density uniformly beyond the SOI.
If the SNR survives to these large radii, we find that they eventually expand nearly spherically, and that the shape in this late stage is largely indifferent to the processes that occurred near the SMBH.
With a uniform density and approximately spherical evolution, the late-time kinematics closely follow the Sedov--Taylor solution.
For densities that flatten outside the SOI to about $n \approx 1~\mathrm{cm}^{-3}$, we find that the SNRs transition to the radiative stage at $\sim 10^4~\mathrm{yr}$, independently of $\Mbh$.
All adiabatic lifetimes therefore end at this age if the SNR has not already been destroyed by shearing from the SMBH.

\subsubsection{Mean lifetimes} \label{sec:results_mean_lifetimes}
Deriving properties over the whole SOI is useful in the cases where, for more distant galaxies, individual SNRs may not be observationally resolved.
Young SNRs will contribute to the total X-ray emission from these regions, which can be more easily observed. 

To summarize the effect of the SMBH environment on SNRs, we calculate the lifetime for core-collapse SNRs averaged over the entire SOI:
\begin{equation} \label{eq:mean_lifetime}
\langle t_\mathrm{ad} \rangle = \frac{\int_{R_\mathrm{sh}}^{R_\mathrm{SOI}} t_\mathrm{ad}(R) \, n_\mathrm{cc}(R) \, R^2 \, \ud R}{ \int_{R_\mathrm{sh}}^{R_\mathrm{SOI}} n_\mathrm{cc}(R) \, R^2 \, \ud R} ,
\end{equation}
where
\begin{equation}
n_\mathrm{cc}(R) \propto
\begin{dcases*}
\left(\frac{R}{R_\mathrm{b}}\right)^{-2}  & $R \leq R_\mathrm{b}$ \\
\left(\frac{R}{R_\mathrm{b}}\right)^{-4.5} & $R > R_\mathrm{b}$
\end{dcases*}
\label{eq:massive_star_density}
\end{equation}
is the volume number density of stars with mass over $8 M_{\odot}$ (see Section~\ref{sec:method_massive_star_distributions}).

The average lifetime, $\langle t_\mathrm{ad} \rangle$ is shown in Fig.~\ref{fig:lifetime_means}, as a function of $\Mbh$.
Examining Fig.~\ref{fig:lifetimes} and equation (\ref{eq:massive_star_density}), we see that much of the reduction in $\langle t_\mathrm{ad} \rangle$ is determined by the value of $t_\mathrm{ad}$ for explosions near and inside the break in gas density ($a \lesssim R_\mathrm{b}$).
The weighted contribution of $t_\mathrm{ad}$ to the mean lifetime is higher inside the break due to the higher density\footnote{Note that $\ud N (M>8) = n_\mathrm{cc}(R)R^2 \, \ud R$ is constant for $R \leq R_\mathrm{b}$.} of massive stars; additionally, much of the reduction in $t_\mathrm{ad}$ occurs near the break for lower values of $\Mbh$, while $t_\mathrm{ad}$ is low throughout the inner region for higher $\Mbh$.
For $\Mbh > 10^7 \Msun$, the mean adiabatic lifetime of the SNR gets increasingly shorter, well below the canonical value of $\sim 10^4~\mathrm{yr}$.
By $\Mbh \gtrsim 10^8 \Msun$, the lifetime of most SNRs in the SOI is ended by disruption by the SMBH while the SNR is in the adiabatic stage.
Shallower inner gas density profiles (green and red lines) amplify these trends.
\begin{figure}
\begin{center}
\includegraphics[width=\columnwidth]{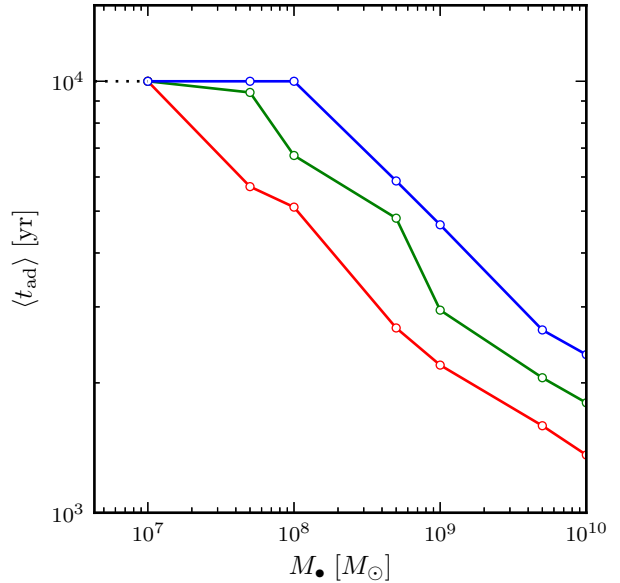}
\caption{Estimated mean adiabatic lifetimes as a function of black hole mass in the present models, using equation (\ref{eq:mean_lifetime}) with the results for $t(R)$ in Fig.~\ref{fig:lifetimes} and the stellar number density distributions $n(R)$ in Section~\ref{sec:method_massive_star_distributions}.
The red (lowermost) line is for $\omega_\mathrm{in} = 1/2$, green for $\omega_\mathrm{in} = 1$ and blue (uppermost) for $\omega_\mathrm{in} = 3/2$.
For lower $\Mbh$ ($\lesssim 10^7~\Msun$), remnants throughout the SOI are not destroyed by the shearing condition by the time radiative losses start to become significant.
The adiabatic lifetime therefore ends at $10^4~\mathrm{yr}$ due to the radiative stage, which is indicated by the dotted black line.}
\label{fig:lifetime_means}
\end{center}
\end{figure}

\section{Discussion and Conclusions} \label{sec:conclusion}
In this paper, we presented a novel numerical method based on the Kompaneets approximation for calculating the evolution of a shock in an arbitrary axisymmetric configuration of density.
Our approach has the benefit of being more flexible than analytic solutions (which only exist for some simple density stratifications) while being much faster than full hydrodynamical simulations.

We apply this numerical method to trace the evolution of SNRs in quiescent galactic nuclei, with properties similar to those of our Galactic Centre.
We describe these nuclei as self-regulating and steady-state systems, where gas inflow from outside this region has been limited and unimportant during at least the last $100~\mathrm{Myr}$.
In this scenario, most of the star formation occurs in situ, recycling the gas ejected in supernovae, and winds from massive stars feed the accretion flow, which forms most of the interstellar gas.
We predict the morphological evolution of SNRs in these nuclei, and relevant time-scales such as their X-ray lifetimes.

We find that the supernova remnants that explode very near low-mass SMBHs ($ \lesssim 10^8~\Msun$, such as Sgr A*) will pass over the SMBH before appreciably decelerating and will continue expanding almost spherically.
Although there can be prominent distortion during the early evolution of SNRs that explode further away from these SMBHs, SNRs at all explosion distances appear reasonably spherical by the onset of the radiative regime, much like SNRs in a typical interstellar medium.
Notably, in the Galactic Centre, this implies that an SNR should be observable in X-ray for $\sim 10^4~\mathrm{yr}$.
The presence of a suspected SNR enveloping Sgr A*, known as Sgr A East \citep{Maeda02}, fits with our prediction of SNRs being able to survive their adiabatic expansion through the Galactic Centre region.

If the SMBH mass is large enough ($\gtrsim 10^8~\Msun$), we instead expect a wide range of SNR morphologies, depending on the explosion distance from the SMBH.
There, SNRs typically end their life due to tidal shearing and disruption.
The observable lifetime is therefore suppressed ($10^2 \sim 10^3~\mathrm{yr}$) with respect to SNRs evolving around lower mass SMBHs.
The reductions in SNR lifetime depend on the inner gradients of gas density as predicted by accretion theory.
Conversely, therefore, observations of SNRs can be used to infer and constrain properties of their environment.
For example, variations in the density gradients can produce different global quantities such as the mean lifetime of SNRs (Fig.~\ref{fig:lifetime_means}) or their overall sizes in their early evolution (Figs~\ref{fig:deceleration_lengths_MW} and \ref{fig:deceleration_lengths}).

The disruption of the SNR, or fraction of it, by the central SMBH, that takes place when this material slows down below the Keplerian velocity would lead to a period of enhanced accretion on to the central black hole.
If a significant fraction of the sheared material is trapped by the SMBH we expect accretion of a few solar masses on to the SMBH over a period of $\sim 100~\mathrm{yr}$, yielding (assuming efficiency of $0.1$) a luminosity of the order of $0.7 \times 10^{44}~\mathrm{erg} \, \mathrm{s}^{-1}$.
This enhanced accretion would be at the sub-Eddington rate for the higher mass black holes (above $0.5 \times 10^6~\Msun$) but still very significant and at a level comparable to a powerful AGN.
This may lead to a period of flaring of the otherwise quiescent black hole.
Such events would happen even around the Galactic Center and other small-mass SMBHs. 
While SNRs are not completely disrupted around such black holes we still expect events in which up to $20\%$ of the SNR material is accreted on to the central black hole over a period of about a hundred years leading to significant flaring.

Beyond the Milky Way, an excellent example of an SNR that is resolved in a galactic nucleus is S Andromedae (SN 1885A), which has an angular diameter of about 0.7 arcsec and has a morphology resolvable by the \textit{Hubble Space Telescope} \citep{Chavalier88, Fesen99, Fesen07}.
The SNR is only 60 parsecs from the centre of the bulge of the Andromeda galaxy, though not quite within the SMBH SOI.
Although SNRs such as S Andromedae are resolvable in other galactic nuclei with the current generation of instruments, individual SNRs may not be distinguished in more distant galaxies.

For those distant galaxies, it is possible to use our formalism to predict global quantities that can be observed, such as the number of SNRs expected at a given time and therefore their total X-ray luminosity (Rimoldi et al., in preparation).
Exploiting the link between SNRs and young massive stars, it is also possible to estimate the expected SFR in the spheres of influence of quiescent SMBHs.
These studies, which will be presented in follow-up work, and their comparison with observation, can inform theory of nuclear assembly and galaxy formation in general.

\section*{Acknowledgements}
This work was supported by the Netherlands Research Council (NWO grant numbers 612.071.305 [LGM] and 639.073.803 [VICI]) and by the Netherlands Research School for Astronomy (NOVA). TP was supported by the ISF I-Core centre for excellence and by a grant from the Israel Space Agency (ISA). We thank an anonymous referee for helpful comments.

\appendix
\section{Integrals of density in the ejecta-dominated stage} \label{ap:integrals}
Here, we elaborate on the treatment of the integrals over the density discussed in Section~\ref{sec:method_deceleration_lengths}.
The general angle-dependent approach is given, as well as the integrations over a sphere for reference.

\subsection{Angle-dependent integrals}
Beginning with equation (\ref{eq:solid_angle_integral}), we consider explosions either outside or inside the break in density.
In general, the integrals along $R'$ can be partitioned into segments between some radii $R'_\mathrm{lower}$ and $R'_\mathrm{upper}$,
\begin{equation} \label{eq:solid_angle_part_integral}
M_{\mathrm{part}} = \rho_a a^\omega \int_{R'_\mathrm{lower}}^{R'_\mathrm{upper}} R^{-\omega} R'^2 \, \ud R' ,
\end{equation}
which, when summed to compose the full integral from $R' = 0$ to $R' = L$, give the full mass swept out over the path.
The integrals are split in such a way to calculate sections that are entirely within one of the two possible density gradients $\omega$.
 
For explosions {outside} the break in density, if a shock segment has an initial angle $\psi_0 > \sin^{-1}{\left( R_\mathrm{b}/a \right)}$, the segment will not cross into the region interior to the break, and the integral is fully through the $\omega = 3$ region.
However, if $\psi_0 \leq \sin^{-1}{\left( R_\mathrm{b}/a \right)}$, the element of the shock crosses the break.
In such cases, the radius $R'$ extending from $R = a$ has either one or two\footnote{For $\psi_0 = \sin^{-1}{\left(R_\mathrm{b}/a \right)}$ and $\psi_0 < \sin^{-1}{\left( R_\mathrm{b}/a \right)}$, respectively} solutions for the intersection with the sphere of radius (measured from the SMBH) equal to $R_\mathrm{b}$:
\begin{equation} \label{eq:R_pm}
R'_{\mathrm{b}\pm} = \frac{1}{2} \left(s \pm \sqrt{s^2 - 4 \left( a^2 - R_\mathrm{b}^2 \right)} \right) ,
\end{equation}
where $s \equiv 2a\cos{\psi_0}$.
For explosions outside the break, the integral can be split into three possible regions (with $R'_{\mathrm{b}\pm}$ provided by equation \ref{eq:R_pm}): $R' < R'_{\mathrm{b}-}$, or $R'_{\mathrm{b}-} < R' < R'_{\mathrm{b}+}$, or $R' > R'_{\mathrm{b}+}$.
These ranges define the integral limits, where each integral has the general form
\begin{equation} \label{eq:r_solidangle}
M_{\mathrm{part}} \propto a^\omega \int {\frac{r^2}{\left(a^2 + R'^2 - s R'\right)^{\omega/2}}} \, \ud R' .
\end{equation}

The same holds for explosions {inside} the break, except that there is only one solution, $R'_\mathrm{b}$ for the intersection with the surface at $R = R_\mathrm{b}$ and equation (\ref{eq:r_solidangle}) has only two sets of limits: $R' < R'_\mathrm{b}$, and $R' > R'_\mathrm{b}$.
The general solutions to the angle-dependent integrals using these limits are lengthy and are not reproduced here. 

\subsection{Integrals over a sphere}
Due to the introduction of an axis of symmetry by the offset position of the sphere, the mass integral may be simplified using cylindrical coordinates with this symmetry axis (the cylindrical $z$) aligned on the explosion point.
A spherically symmetric density field (having an origin coincident with that of this cylindrical coordinate system) remains constant over the cylindrical polar angle $\varphi$, for a given cylindrical\footnote{Note that we use $R$ to designate the spherical radial coordinate, and $r$ to designate the cylindrical one.} $r$ and $z$, since the spherical $R$ ($= \sqrt{r^2 + z^2}$) is constant.
Thus the expression for mass is reduced to a double integral.

In such a coordinate system, for a {single} power-law density of exponent $-\omega$, the mass swept up ($M_\mathrm{s}$) by a spherical shock front that has expanded through a radial distance $L$ (measured from the explosion point, $a$)
\begin{equation} \label{eq:mswept}
M_\mathrm{s} = 2 \uppi \rho_a a^{\omega} \int_{a - L}^{a + L} \int_0^{\sqrt{L^2 - (z - a)^2}} r \left(r^2 + z^2\right)^{-\omega/2} \, \ud r \, \ud z .
\end{equation}
For the single density distributions of $\omega = 1$ and $3$, this evaluates to
\begin{equation}
M_\mathrm{s} =
\begin{dcases*}
\frac{2 \uppi \rho_a}{3} \left[\left(a + L\right)^3 - \abs{a - L}^3 - 6 a^2 L\right] , & $\omega = 1$ \\
2 \uppi \rho_a a^3 \left[\ln{\left(\frac{a + L}{a - L}\right)} - \frac{2L}{a}\right] , \; a > L, & $\omega = 3$ .
\end{dcases*}
\end{equation}
In the case $a > L$ for the $\omega = 1$ solution, the term in brackets reduces to $2 L^3$ such that $M_\mathrm{s} = 4 \uppi \rho_0 L^3/3$, which is the trivial $\omega = 0$ solution.

For a model density in which there is a {broken} power-law distribution, the integral over a sphere centred at $z=a$ is less straightforward.
To avoid introducing complicated integral limits to equation (\ref{eq:mswept}), one approach is to determine the overall quantity by summing integrals over two density distributions, where the integrand for each is restricted using Heaviside step functions, $H$, that break the distribution at specified (spherical) radii, $R$.
Using the cylindrical integral of equation (\ref{eq:mswept}), and taking a density distribution that is non-zero between two spherical radii from the origin, $R = P$ and $R = Q$, this effectively constrains the density as
\begin{equation}
\rho \rightarrow \rho \times \left[H \left(\sqrt{r^2 + z^2} - P\right) - H \left(\sqrt{r^2 + z^2} - Q\right)\right] .
\end{equation}
For a two-section broken power-law density with $\omega_\mathrm{in}$ and $\omega_\mathrm{out}$, the total swept-up mass will be $M_\mathrm{s} = M_\mathrm{in} + M_\mathrm{out}$.
The solutions for $\omega = 1$ and $3$ are given below.

As an alternative method, solutions for $\omega = 1/2$ and $3/2$ were obtained using integrals of the unconstrained density by instead splitting the integrated regions into spherical caps, and adjusting the integral limits appropriately.
These curves match the behaviour of those for $\omega = 1$ and $3$, showing agreement between the two methods of integration.

\subsubsection{Solution for $\omega=1$}
For an explosion at $R = a$ and an outer break at $R = c$:
\begin{multline}
M_1 = \frac{\uppi \rho_0}{3} \left[H\left(a + L\right) \left\{\left(a + L\right)^2 \left[2 \left(a + L\right) - 3a\right]\right\}\right. \\
      - H\left(a + L - c\right) \left\{\left(a + L - c\right)^2 \left[2 \left(a + L \right) + c - 3a\right]\right\} \\
      - H\left(\abs{a - L}\right) \left\{\left(a - L\right)^2 \left[2 \abs{a-L} - 3a \right] \right\} \\
      + \left. H\left(\abs{a - L} - c\right) \left\{\left(\abs{a - L} - c\right)^2 \left[2 \abs{a - L} + c - 3a\right]\right\}\right] .
\end{multline}

\subsubsection{Solution for $\omega=3$}
In this case, an inner density break at $R=c$ needs to be applied:
\begin{multline}
M_3 = 2 \uppi \rho_0 a^3 \left[H\left(a + L - c\right) \left\{\frac{\left(c - a - L\right)^2}{2ac} \right.\right. \\
      \left. - \ln{\left(\frac{c}{a+L}\right)} - \frac{a + L}{c} + 1 \right\} \\
      + H\left(a - L - c\right) \left\{\ln{\left(\frac{c}{a - L}\right)} + \frac{a - L}{c} - 1 \right\} \\
      \left. + H\left(\abs{a - L} - c\right) \left\{\frac{\left(c - \abs{a-L}\right)^2}{2ac} \right\}\right] .
\end{multline}

\section{Analytic volume, time and velocity expressions in the Kompaneets approximation}
\label{ap:kompaneets_solutions}
Solutions for the volume, time and velocity are given in K92, with the given coordinate transformation, for power laws of $\omega = 0$ and $4$.
Instead, $\omega = 1$ and $3$ are relevant for the presently considered density profiles, and we outline here the corresponding solutions.

In a general radial power-law profile, the solution for the volume is (see equation \ref{eq:Psi}, as well as equation 16 in K92)
\begin{align}
V &= 2 \uppi \int_{R_-}^{R_+} {\left(1 - \cos{\theta}\right) R^2 \, \ud R} \notag \\
  &= 2 \uppi \int_{R_-}^{R_+} {\left\{1 -\cos{\left(\frac{1}{\alpha} \cos^{-1}{\left[\frac{1 - x^2 + \left(\frac{R}{a}\right)^{2\alpha}}{2\left(\frac{R}{a}\right)^{\alpha}}\right]}\right)}\right\} R^2} \, \ud R , \notag \\
\end{align}
recalling that $\alpha \equiv \left(2 - \omega\right)/2$. For $\omega = 1$ and $3$,
\begin{equation}
R_\pm = 
\begin{dcases*}
a \left(1 \pm x\right)^{2}, & $\omega = 1$ \\
\frac{a}{ \left(1 \mp x\right)^{2}}, & $\omega = 3$
\end{dcases*}
\end{equation}
are the leading ($R_+$) and trailing ($R_-$) points of the shock (along $\theta = 0$, using equation \ref{eq:Psi}).
When comparing the equations for $\omega = 1$ and $3$ with the $\psi_0$ parametrisation, it is important to note that a phase shift of $\uppi$ in $\psi_0$ is required to obtain correspondence (of the definition of $R_-$ and $R_+$, for example) between the two cases.
In terms of $x$, this gives the following solutions:\footnote{The identity $\cos{\left(2\cos^{-1}z\right)} = 2z^2 - 1$ is useful for the solution to these integrals.}
\begin{align}
V(x) = \kappa_V \times
\begin{dcases*} 
x^3 \left(1 + x^2\right), & $\omega = 1$ \\
x^3 \frac{\left(1 + x^2\right)}{\left(x^2 - 1\right)^6}, & $\omega = 3$ 
\end{dcases*} 
\end{align}
for
\begin{align}
\kappa_V \equiv \frac{32 \uppi a^3}{3} .
\end{align}
Therefore the time can be found from the following integrals of equation (\ref{eq:dydt}):
\begin{align}
t(x) = \kappa_t \times
\begin{dcases*} 
\int_0^x{\sqrt{u^3 \left(1 + u^2\right)}} \, \ud u , & $\omega = 1$ \\
\int_0^x{\sqrt{\frac{u^3 \left(1 + u^2\right)}{\left(u^2 - 1\right)^6}}} \, \ud u , & $\omega = 3$ 
\end{dcases*}
\end{align}
for
\begin{equation}
\kappa_t \equiv \sqrt{\frac{256 \uppi a^5 \rho_0}{3 \lambda E \left(\gamma^2 - 1\right)}} .
\end{equation}
The $\omega = 1$ result may be written via a power-series expansion about $x=0$,
\begin{equation}
t_1(x) = \kappa_t \left( \frac{2x^{5/2}}{5} + \frac{x^{9/2}}{9} - \frac{x^{13/2}}{52} + \ldots \right) .
\end{equation}
The integral for $\omega = 3$ is divergent for $x \rightarrow 1$, and is represented by the power series about $x=0$
\begin{equation}
t_3(x) = \kappa_t \left(\frac{2x^{5/2}}{5} + \frac{x^{9/2}}{9} + \frac{59 x^{13/2}}{52} + \ldots \right) .
\end{equation}

For $\omega = 1$, the value of $x = 1 \leftrightarrow y = y_c$ signifies the moment at which the trailing point of the shock reaches the density singularity ($R_- = 0$).
This is, therefore, also the onset of shock self-interactions as other parts of the shock wrap around this point.
This occurs in a finite time $t$, and, unlike in the $\omega = 3$ case, solutions still exist beyond $x = 1$ (there is no blow-out of the shock front, as in the case of $\omega = 3$ where the leading point $R_+ \rightarrow \infty$).

In terms of $x$, we therefore have
\begin{equation}
v_\mathrm{n}(x) = \kappa_v \times
\begin{dcases*}
\sqrt{\frac{1}{ x^3 \left( 1 + x^2 \right) } }, & $\omega = 1$\\
\sqrt{\frac{ \left(x^2 - 1\right)^6}{ x^3 \left( 1 + x^2 \right) } }, & $\omega = 3$
\end{dcases*}
\end{equation}
for
\begin{equation}
\kappa_v \equiv \sqrt{\frac{3 \lambda E \left(\gamma^2 - 1\right)}{64 \uppi a^3 \rho(R)}} .
\end{equation}
Together, these expressions allow a transformation of the solution in terms of $x$ into physical units.
We are now in a position to compare the numerical solutions with analytic ones, as is done in Section~\ref{sec:method_compare_analytic_numerical}.

\section{Numerical treatment of shock self-interactions}\label{ap:self_interactions}
Here, we outline the treatments of self-interacting segments of the shock front in our numerical scheme.
In an axisymmetric arrangement, self-interactions can happen in two ways.
In the first case, any parts of the shock that pass over the axis of symmetry (the $x$-axis in our coordinates) will collide with the complementary part of the shock travelling over the axis in the opposite direction.
This can happen as the shock wraps around the SMBH, where the density is at its highest.
Other shock self-interactions can be caused by variations in the density profile that force flowlines to converge, such as near a break in power-law densities (see Fig.~\ref{fig:flowline_mergers} for an example).

We therefore need a routine that can detect self-interactions in a general way.
A simple implementation would be to examine the location histories of the flowlines to determine if any have intersected.
In our experience, storing these histories puts too high a demand on memory.
Instead, two methods for detecting interacting regions in the shock front were investigated.

The first approach is an anticipatory one, which tests the spatial divergence of the velocity of the shock front at each time step.
This quantity gives an indication of whether portions of the shock front are converging.
However, care is required in choosing the threshold of divergence used to define merging regions of the shock, which results in this divergence method being difficult to tune.

In practice, it is more straightforward to use a reactive detection of self-interactions.
The shock front, defined by the positions of the flowlines at one point in time, is constrained algorithmically to be a simple piecewise linear curve.
Any intersections along the curve can be detected, as we monitor the ordering of the points.
Loops arising from intersections of the shock front are removed, which is equivalent to an effective merger of all flowlines involved in the intersecting loop into a single resultant flowline.

In all cases of self-interactions, the merged flowlines are replaced by a single flowline with an average of their positions.
The ejecta mass represented by the new flowline is taken to be the sum of the masses assigned to the previously merging flowlines.

Fig.~\ref{fig:flowline_mergers} shows an example of regions along a sample shock solution in which flowlines are converging and being merged.
This specific example models fluid elements (flowlines; dashed curves) at the shock front (solid lines) colliding due to the change in gradient of the background density (thick blue line).
\begin{figure}
\begin{center}
\includegraphics[width=\columnwidth]{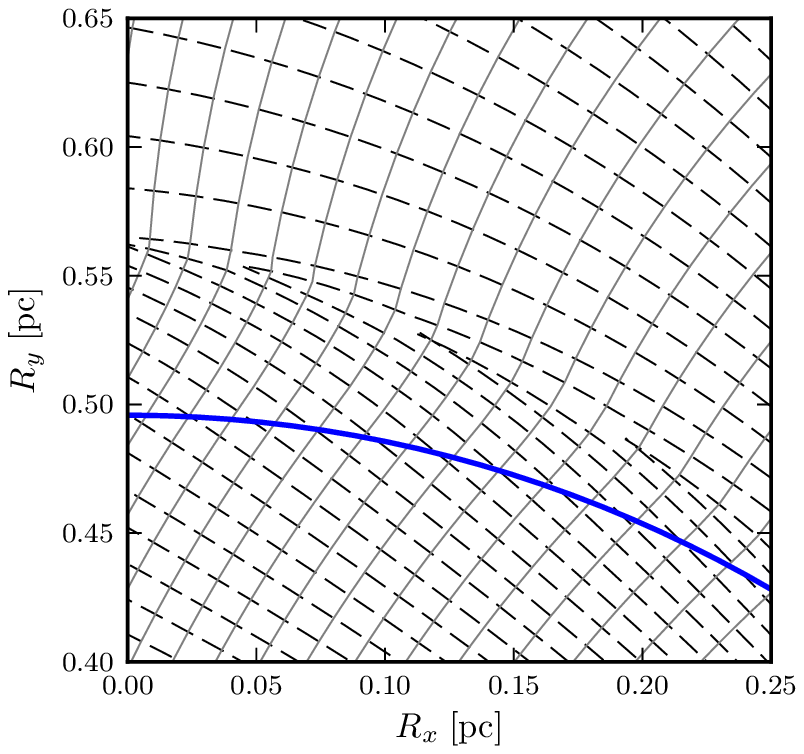}
\caption{Example of flowline intersection and merging due to the presence of a boundary (thick blue line) between two power-law densities.
Dashed lines show flowline paths, which lie perpendicular to the shock front (sample shock snapshots are given as solid grey lines).
In this example, the break is between power laws of $\omega_\mathrm{in} = 1/2$ and $\omega_\mathrm{out} = 3$, for an explosion around a $10^7~\Msun$ SMBH at $R_x = 0.6~\mathrm{pc}$, $R_y = 0$.}
\label{fig:flowline_mergers}
\end{center}
\end{figure}

\footnotesize{

}

\label{lastpage}
\end{document}